%% file: acl_latex.tex
\pdfoutput=1

\documentclass[11pt]{article}
\PassOptionsToPackage{table}{xcolor}  
\usepackage{xcolor} 
\usepackage[final]{acl}

\usepackage{times}
\usepackage{latexsym}

\usepackage{bm}

\usepackage{tcolorbox}
\usepackage{wrapfig}

\usepackage{amsmath}  
\usepackage{amssymb}  
\usepackage{amsfonts}

\usepackage{booktabs}
\usepackage{tikz}
\usetikzlibrary{shapes.geometric}
\usepackage{multirow}
\usepackage{array}
\usepackage{colortbl}

\usepackage{subfigure}
\usepackage{booktabs}
\usepackage{tabularx}
\usepackage{makecell}
\usepackage{caption}
\usepackage{xurl}
\usepackage[normalem]{ulem}
\usepackage{soul}
\usepackage{array}
\usepackage{geometry}

\usepackage{tikz}
\usepackage{colortbl}
\usepackage{algorithm}
\usepackage{algorithmic}
\usepackage{enumitem}
\usepackage{stfloats}

\definecolor{lightblue}{RGB}{221, 231, 245} 
\definecolor{lightyellow}{RGB}{209, 239, 241}
\definecolor{lightgreen}{RGB}{255, 240, 230}
\definecolor{lightred}{RGB}{255,102,102}

\usepackage[T1]{fontenc}

\usepackage[utf8]{inputenc}

\usepackage{microtype}

\usepackage{inconsolata}

\usepackage{graphicx}

\def\method{ExpandR}

\title{\method{}: Teaching Dense Retrievers Beyond Queries with LLM Guidance}

\author{Sijia Yao$^{1}$\thanks{ \ \ indicates equal contribution.}, Pengcheng Huang$^{1}$\footnotemark[1], Zhenghao Liu$^{1}$\thanks{ \ \ indicates corresponding author.}, Yu Gu$^{1}$,\\ 
\textbf{  Yukun Yan$^{2}$, Shi Yu$^{2}$, Ge Yu$^{1}$} \\
$^1$Department of Computer Science and Technology, Northeastern University, China \\
$^2$Department of Computer Science and Technology, Institute for AI, Tsinghua University, China \\
    Beijing National Research Center for Information Science and Technology, China 
}

\begin{document}
\maketitle

\input{section/0_abstract}

\input{section/1_introduction}

\input{section/2_related-work}

\input{section/3_method}

\input{section/4_experiment}
\input{section/5_result}

\input{section/6_conclusion}
\input{section/7_limitation}

\bibliography{acl_latex}
\clearpage
\appendix
\input{section/8_appendix}

\end{document}

%% file: section/0_abstract.tex
\begin{abstract}
Large language models (LLMs) have demonstrated significant potential in enhancing dense retrieval through query augmentation. However, most existing methods treat the LLM and the retriever as separate modules, overlooking the alignment between generation and ranking objectives. In this work, we propose \textbf{\method{}}, a unified LLM-augmented dense retrieval framework that jointly optimizes both the LLM and the retriever.  \method{} employs the LLM to generate semantically rich query expansions, which are leveraged to enhance the retriever's training. Simultaneously, the LLM is trained using Direct Preference Optimization (DPO), guided by a carefully designed reward function that balances retrieval effectiveness and generation consistency. This joint optimization paradigm enables mutual adaptation between the LLM and the retriever, resulting in query expansions that are both informative and well-suited for retrieval. Experimental results on multiple benchmarks show that \method{} consistently outperforms strong baselines, achieving more than a 5\% improvement in retrieval performance. All codes are available at \url{https://github.com/NEUIR/ExpandR}.

\end{abstract}

%% file: section/1_introduction.tex
\section{Introduction}
Dense retrievers~\cite{karpukhin2020dense,xiong2021approximate} encode both queries and documents into the same embedding space, enabling efficient similarity-based retrieval via approximate KNN search~\cite{johnson2019billion}.
While effective, their performance remains highly sensitive to the quality of the input query. In practice, user queries~\cite{belkin1982ask,ingwersen1996cognitive} are often short and ambiguous, leading to a significant semantic gap between the query and relevant documents, making it challenging for dense retrievers to accurately capture the underlying information need.

Recent advances in Large Language Models (LLMs) offer promising solutions to this challenge through query augmentation~\cite{wei2022chain,huang2024translate,wei2022emergent}. Existing methods along this line of research can be categorized into two groups.
The first direction leverages LLM-generated reformulations as supervision signals to train dense retrieval models, typically through contrastive training~\cite{zhang2025unleashing, ma2025drama} or ranking probability distillation~\cite{shi2024replug,kim2025syntriever}. However, the effectiveness of this approach is constrained by the limited capacity and scalability of dense retrievers~\cite{fang2024scaling}.
The second direction focuses on augmenting dense retrievers by prompting LLMs to generate additional terms at inference time~\cite{wang2023query2doc, mackie2023generative}. These terms aim to increase lexical overlap with relevant documents, thereby reducing the semantic gap between queries and documents. While such expansions are often semantically rich, they are typically misaligned with the retriever, as the LLM is not explicitly optimized for retrieval objectives. As a result, the retriever struggles to effectively utilize the LLM-augmented content.

In this work, we propose \textbf{\method{}}, a unified LLM-augmented dense retrieval framework that jointly optimizes both the LLM and the dense retriever. \method{} first prompts the LLM to generate semantically enriched query expansions, which enhance query representations and improve the retriever's ability to rank relevant documents. Rather than treating the LLM and retriever as separate modules, \method{} integrates generation and retrieval under a shared training objective---promoting higher ranks for ground-truth documents given a query.
Specifically, we optimize the dense retriever via contrastive training, and train the LLM using Direct Preference Optimization (DPO) with a combination of self-consistency and retrieval-based rewards.
Through this joint optimization, the two components mutually reinforce each other, leading to more effective expansions and improved overall retrieval performance.

Our experiments on the BEIR benchmark~\cite{thakur2021beir} demonstrate the effectiveness of \method{}, yielding over a 5.8\% improvement in supervised dense retrieval. Further analysis shows that the query expansions generated by \method{} lead to better alignment with relevant documents compared to those from baseline methods. By jointly leveraging self-consistency and retrieval-based rewards, the LLM is better optimized to generate expansions that are both semantically rich and retriever-aligned. Specifically, the self-consistency reward encourages the LLM to generate content that is semantically closer to the ground-truth document, while the retrieval-based reward captures the retriever's ranking behavior. Together, these rewards guide the LLM to produce expansions that are both relevant and retriever-friendly.

%% file: section/2_related-work.tex
\section{Related Work}

Dense retrievers~\cite{karpukhin2020dense,xiong2021approximate,izacard2021unsupervised,Yu2021FewShotCD,xiong2021dense,li2021more} conduct semantic matching by encoding queries and documents into a shared embedding space, thereby alleviating the vocabulary mismatch problem~\cite{belkin1982ask}. To further improve the quality of semantic matching, recent work has focused on refining this embedding space through contrastive learning with relevance supervision~\cite{karpukhin2020dense,zhan2021optimizing} or leveraging weakly supervised training signals~\cite{xie2023unsupervised}. While effective, a persistent bottleneck in information retrieval lies in the quality of the user-issued queries themselves~\cite{jiang2025deepretrieval}. In particular, queries are often underspecified, ambiguous, or semantically incomplete, which limits the retriever's ability to accurately locate relevant content~\cite{belkin1982ask,ingwersen1996cognitive}.

Recent advances in LLMs~\cite{achiam2023gpt, glm2024chatglm} offer new opportunities to address this issue by leveraging their rich knowledge and powerful generative capabilities to enrich or reformulate user queries~\cite{yu2020few, lin2020conversational, ye2023enhancing}. These augmented queries are often used as supervision signals or distillation targets to train dense retrievers more effectively. For instance, methods such as LLM-QL~\cite{zhang2025unleashing} and DRAMA~\cite{ma2025drama} propose leveraging LLMs to generate new queries or training triplets for dense retriever optimization. RePlug~\cite{shi2024replug} has been proposed to distill the knowledge of LLMs into a lightweight retriever. While these approaches enhance supervised retrieval performance, they mainly focus on query synthesis, often overlooking the limited semantic expressiveness of the original queries~\cite{wang2023generative}. Moreover, their effectiveness is fundamentally constrained by the limited capacity and scalability of dense retrievers~\cite{huang2024scalingnote}.

LLM-based query expansion has emerged as a widely adopted approach for query augmentation, effectively enriching the semantic content of original queries. These methods prompt LLMs to generate query-related documents~\cite{wang2023query2doc, jagerman2023query, gao2023precise}, leverage Chain-of-Thought (CoT) reasoning results~\cite{wei2022chain, trivedi2023interleaving}, or utilize specific keywords~\cite{li2024can, jagerman2023query} to expand queries, thereby enhancing the ranking capabilities of lexical matching based retrieval models~\cite{jagerman2023query, wang2023query2doc}, dense retrieval models~\cite{wang2023query2doc}, and reranking models~\cite{li2024can}. However, these LLM-generated expansions are often directly incorporated into the retrieval process without retraining or adapting the retriever. Consequently, the retriever fails to fully leverage the enriched signals of LLMs, resulting in limited improvements in retrieval performance~\cite{wang2023query2doc}.

Moreover, existing approaches that incorporate LLMs into retrieval systems often train the LLM~\cite{jiang2025deepretrieval} or the retriever independently~\cite{kim2025syntriever}, resulting in preference misalignment between the generation and retrieval components. Some works, such as RaFe~\cite{mao2024rafe}, attempt to align LLM rewriting with retrieval signals by using reranker scores as feedback. However, these approaches rely on a separate reranking model rather than incorporating direct training signals from dense retrievers. In contrast, our approach introduces a joint training framework that simultaneously optimizes the LLM and the dense retriever, enabling stronger alignment between the two components to conduct a more effective retrieval result.

%% file: section/3_method.tex
\input{Figure/model}
\section{\method{}: An LLM Augmented Dense Retriever Method}

As illustrated in Figure~\ref{fig:model_pipline}, this section introduces \method{}, our LLM-augmented dense retrieval model that leverages query expansions to improve retrieval performance. We begin by describing the overall architecture of \method{} (Sec.~\ref{model:joint-training}). We then present how LLM-generated query expansions are used to guide the training of the dense retriever (Sec.~\ref{model:contrastive-training}). Finally, we detail a preference-based optimization strategy for the LLM to generate more effective and tailored query expansions (Sec.~\ref{model:llm-optimization}).


\subsection{Toward a Framework for LLM-Guided Dense Retrieval}\label{model:joint-training}
This section illustrates how LLMs can be leveraged to enhance dense retrieval. We first introduce the architecture of a standard dense retriever, and then present our proposed method, \method{}, which incorporates LLM guidance to improve the retrieval model's effectiveness.

\textbf{Dense Retrieval.} Given a query $q$ and a document collection $\mathcal{D} = \{d_\text{1}, ..., d_\text{k}\}$, dense retrieval models~\cite{karpukhin2020dense,xiong2021approximate,cocondenser} first encode the query $q$ and the $i$-th document $d_i$ into embeddings $\vec{q}$ and $\vec{d}_i$ using PLMs, such as BERT~\cite{devlin2019bert}:
\begin{equation}\label{eq:encode}
\small
    \vec{q} = \text{BERT}_q (q),  \quad  \vec{d}_i = \text{BERT}_d (d_i).
\end{equation}
Then the relevance score $S(q, d_i)$ is calculated to estimate the relevance between $q$ and $d_i$:
\begin{equation}
\small
    S(q, d_i) = \text{sim} (\vec{q}, \vec{d}_i),
\end{equation}
where sim is the dot product operation. Finally, dense retrieval models conduct a KNN search~\cite{douze2024faiss} to retrieve the top-ranked documents to satisfy the user needs.

\textbf{\method{}.} Unlike traditional dense retrieval models~\cite{karpukhin2020dense}, \method{} leverages the knowledge encoded in LLM to guide dense retrievers via query expansions $d^{\text{exp}}$, aiming to achieve more accurate retrieval results.

Specifically, we first prompt the LLM $\mathcal{M}$ to generate a query expansion as follows:
\begin{equation}
\small
d^{\text{exp}} = \mathcal{M}(\text{Instruct}_\text{q2d}, q),
\end{equation}
where $\text{Instruct}_\text{q2d}$ denotes an instruction prompting the LLM to generate an informative expansion for the input query~\cite{jagerman2023query}.
We then model the joint probability of retrieving the ground truth document $d_*$ conditioned on the original query $q$ as:
\begin{equation}
\small
P(d_* \mid q; \Phi, \Theta) = P(d_* \mid q, d^{\text{exp}}; \Phi) \cdot P(d^{\text{exp}} \mid q; \Theta),
\end{equation}
where $\Phi$ and $\Theta$ represent the parameters of the retriever and the LLM, respectively. This formulation can be rewritten as:
\begin{equation}\label{eq:joint_prob}
\small
\begin{aligned}
& \log P(d_* \mid q; \Phi, \Theta) = \\
& \log P(d_* \mid q, d^{\text{exp}}; \Phi) + 
\log P(d^{\text{exp}} \mid q; \Theta).
\end{aligned}
\end{equation}
Our objective is to jointly optimize the retriever and the LLM--i.e., $\Phi$ and $\Theta$--to maximize the above log-likelihood, as described in Section~\ref{model:contrastive-training} and Section~\ref{model:llm-optimization}, respectively.

\subsection{Optimizing Dense Retriever through LLM-Guided Contrastive Training}\label{model:contrastive-training}
To maximize $P(d_* \mid q; \Phi, \Theta)$ by optimizing the retriever parameters $\Phi$, we train the dense retriever using both the original query $q$ and its corresponding expansion $d^{\text{exp}}$:
\begin{equation}
\small
\begin{aligned}
&\log P(d_* \mid q; \Phi, \Theta) \\&= 
\underbrace{\log P(d_* \mid q, d^{\text{exp}}; \Phi)}_{\text{Optimize w.r.t. } \Phi} + 
\underbrace{\log P(d^{\text{exp}} \mid q; \Theta)}_{\text{Fixed}}.
\end{aligned}
\end{equation}
To optimize the retriever, we fix $\Theta$ and update only $\Phi$ by maximizing the retriever-related term:
\begin{equation}
\small
    \Phi^* = \arg\max_{\Phi} \ \log P(d_* \mid q, d^{\text{exp}}; \Phi).
\end{equation}
To incorporate the knowledge of $d^{\text{exp}}$, we simply average the embeddings of both $q$ and $d^{\text{exp}}$ as the final query representation $\vec{q}^{\; \text{exp}}$:
\begin{equation}\label{eq:qexp}
\small
    \vec{q}^{\; \text{exp}} = \frac{\vec{q} + \vec{d}^{\; \text{exp}}}{2}.
\end{equation}
Then we treat the expanded query $q^{\text{exp}}$ as the new query and compute the similarity score $\text{sim}(q^{\text{exp}}, d)$ between $q^{\text{exp}}$ and each candidate document $d$. The retriever can be contrastively trained using the training loss $\mathcal{L}_\text{DR}$:
\begin{equation}
\small
 \mathcal{L}_\text{DR} = -\log\frac{e^{\text{sim}(q^{\text{exp}},d_*)}}
    {e^{\text{sim}(q^{\text{exp}},d_*)} + \sum_{d_-\in \mathcal{D}^-}{e^{\text{sim}(q^{\text{exp}},d_-)}}},
\end{equation}
where $\mathcal{D}^-$ represents the set of negative documents, which are sampled from in-batch negatives~\cite{karpukhin2020dense}. 


\subsection{Optimizing LLM for Aligning with Ranking
Preference}\label{model:llm-optimization}
To maximize the probability $P(d_* \mid q; \Phi, \Theta)$, we optimize only the LLM parameters ($\Theta$) while keeping the dense retriever parameters ($\Phi$) fixed.

As shown in Eq.~\ref{eq:joint_prob}, updating $\Theta$ alone still affects both terms of the joint probability. Therefore, we optimize $\Theta$ as follows:
\begin{equation}\label{eq:reward_modeling}
\small
\begin{aligned}
\Theta^* = \arg\max_{\Theta} [ &\log P(d_* \mid q, d^{\text{exp}}; \Phi) \\
&+ \log P(d^{\text{exp}} \mid q; \Theta) ].
\end{aligned}
\end{equation}
This objective indicates that a well-generated $d^{\text{exp}}$ can not only directly increase the likelihood term $\log P(d^{\text{exp}} \mid q; \Theta)$, but also indirectly improve retrieval performance by providing more informative expansions for the term $\log P(d_* \mid q, d^{\text{exp}}; \Phi)$. To realize this dual effect, we optimize the LLM parameters through a reward-driven approach. The optimization process involves two steps: first, we define the reward modeling objective (Eq.~\ref{eq:reward_modeling}); then, we train the LLM using the Direct Preference Optimization (DPO) method~\cite{amini2024direct}.

\textbf{Reward Modeling.} 
We define a reward function $R(d^{\text{exp}})$ to evaluate each candidate expansion $d^{\text{exp}} \in \mathcal{D}^q$. The reward combines two complementary signals:
\begin{equation}\label{eq:reward}
\small
R(d^{\text{exp}}) = R_{\text{self}}(d^{\text{exp}}) + R_{\text{retriever}}(d^{\text{exp}}),
\end{equation}
where $R_{\text{self}}( d^{\text{exp}})$ and $R_{\text{retriever}}(d^{\text{exp}})$ represent the self-reward and the retriever reward, respectively.

\textit{Self-Reward.} To promote the likelihood term $\log P(d^{\text{exp}} \mid q; \Theta)$, we incorporate a self-reward that leverages the LLM's self-consistency. Specifically, we prompt the LLM to generate an answer $y$ according to the query $q$ and the ground-truth document $d_*$:
\begin{equation}
\small
y = \mathcal{M}(\text{Instruct}_\text{q2a}, q, d_*),
\end{equation}
where $\text{Instruct}_\text{q2a}$ guides the LLM to produce an answer $y$ to $q$. We then treat the answer $y$ as a query and rank the expansion candidates $\mathcal{D}^q$ to compute the self-reward score:
\begin{equation}
\small
R_{\text{self}}(d^{\text{exp}}) = \frac{1}{\text{Rank}(y, d^{\text{exp}})},
\end{equation}
where $\text{Rank}(y, d^{\text{exp}})$ denotes the rank of document $d^{\text{exp}}$ based on its relevance score $\text{sim}(y, d^{\text{exp}})$. A higher rank indicates stronger semantic similarity and consistency between $y$ and $d^{\text{exp}}$.

\textit{Retriever Reward.}
While the self-reward ensures the semantic plausibility of the candidate expansion $d^{\text{exp}}$, it does not necessarily guarantee its usefulness for retrieval, i.e., contributing to $\log P(d_* \mid q, d^{\text{exp}}; \Phi)$~\cite{weller2024generative}.  
To address this limitation, we incorporate a retriever reward that captures the preferences of the retriever.  
Specifically, we compute the Mean Reciprocal Rank (MRR) by treating the ground-truth document $d_*$ as a pseudo-query and ranking the expansion candidates $\mathcal{D}^q$:
\begin{equation}
\small
R_{\text{rank}}(d^{\text{exp}}) = \frac{1}{\text{Rank}(d_*, d^{\text{exp}})},
\end{equation}
where $\text{Rank}(d_*, d^{\text{exp}})$ denotes the rank of $d^{\text{exp}}$ based on the similarity score $\text{sim}(d_*, d^{\text{exp}})$.  
A higher reward indicates that the expansion is more similar to the ground-truth document, and thus more likely to improve retrieval performance.

\textbf{LLM Optimization.} We fine-tune the LLM $\mathcal{M}$ using preference modeling via DPO. Specifically, we first prompt the LLM to generate a set of expansion candidates $\mathcal{D}^q = \{d^{\text{exp}}_1, \dots, d^{\text{exp}}_k\}$ for each query $q$, by sampling with varying temperature:
\begin{equation}
\small
d^{\text{exp}} \sim \mathcal{M}(\text{Instruct}_\text{q2d}, q).
\end{equation}
Then we construct training triples $(q, d^{\text{exp}}_+, d^{\text{exp}}_-)$ using the reward model $R(\cdot)$ (Eq.~\ref{eq:reward}):
\begin{equation}
\small
    R(d^{\text{exp}}_+) > R(d^{\text{exp}}_-),
\end{equation}
and follow the DPO method to optimize the LLM ($\mathcal{M}$) using the loss function $\mathcal{L}(\mathcal{M}; \mathcal{M}^\text{Ref})$:
\begin{equation}
\small 
\begin{aligned}
\label{eq:dpo}
& \mathcal{L}(\mathcal{M}; \mathcal{M}^\text{Ref}) = 
- \mathbb{E}_{(q, d^{\text{exp}}_+,d^{\text{exp}}_-) \sim \mathcal{P}} \Big[ \log \sigma \Big( \\
& \beta \log \frac{\mathcal{M}(d^{\text{exp}}_+ \mid q)}{\mathcal{M}^\text{Ref}(d^{\text{exp}}_+ \mid q)} - 
\beta \log \frac{\mathcal{M}(d^{\text{exp}}_- \mid q)}{\mathcal{M}^\text{Ref}(d^{\text{exp}}_- \mid q)} \Big) \Big],
\end{aligned}
\end{equation}
where $\sigma$ is the sigmoid function, $\beta$ is a scaling hyperparameter, and $\mathcal{M}^{\text{Ref}}$ is a frozen reference model. The training set $\mathcal{P}$ is composed of preference pairs sampled based on reward scores.

%% file: Figure/model.tex
\begin{figure*}[t]
  \centering
  \includegraphics[width=1.0\linewidth]{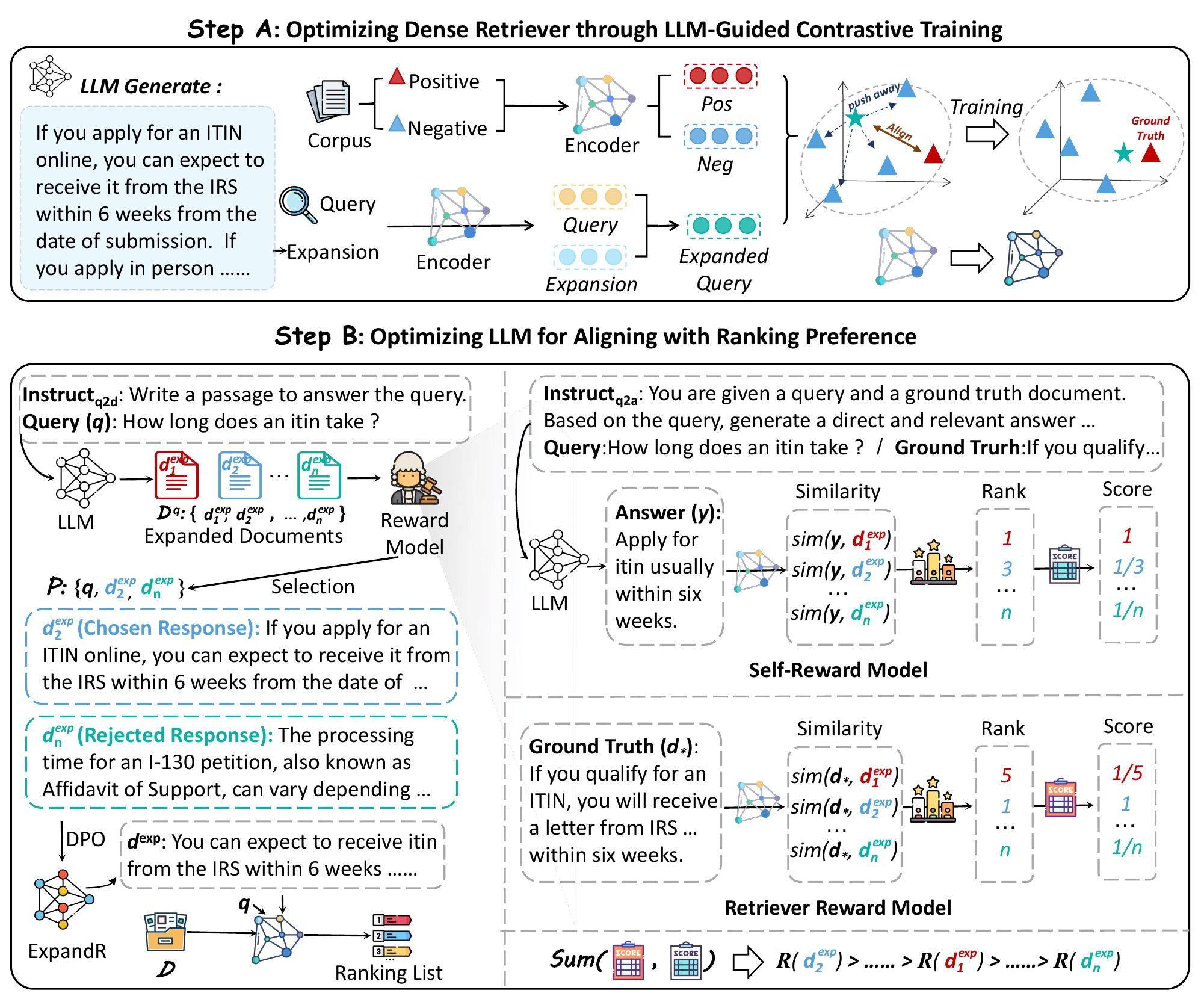}
  \caption{Illustration of Our \method{} Model. \method{} optimizes both dense retriever and LLM using the LLM-guided contrastive training method and the ranking preference alignment method.}
  \label{fig:model_pipline}
\end{figure*}

%% file: section/4_experiment.tex
\section{Experimental Methodology}\label{sec:exp}
In this section, we introduce the datasets, evaluation metrics, baselines, and implementation details used in our experiments. 

\textbf{Dataset.}
We utilize various datasets for training and evaluation. Data statistics are shown in Table~\ref{tab:dataset}. More details on data generation and processing are shown in Appendix~\ref{app:experiment_detail}.

\textit{Training.}
We use the publicly available E5 dataset~\cite{wang2024improving,springer2024repetition} to train both the LLMs and dense retrievers. We concentrate on English-based question answering tasks and collect a total of 808,740 queries. From this set, we randomly sample 100,000 queries to construct the DPO data for training LLM, while the remaining queries are used for contrastively training the dense retrieval model. During the construction of DPO preference pairs, we first prompt LLMs to generate documents as query expansions~\cite{wang2023query2doc}. We then filter out queries whose generated documents exhibit low semantic similarity to the original queries. This results in a final dataset comprising 30,000 high-quality queries.

\textit{Evaluation.}
We evaluate retrieval effectiveness using two retrieval benchmarks: MS MARCO~\cite{bajaj2016ms} and BEIR~\cite{thakur2021beir}.

\textbf{Evaluation Metrics.}
We use nDCG@10 as the evaluation metric, which is the official evaluation metric of BEIR~\cite{thakur2021beir}. Statistical significance is tested using a permutation test with $p<0.05$.


\input{Table/dataset}
\textbf{Baselines.}

We compare our \method{} model with four representative retrieval models, including BM25~\cite{robertson2009probabilistic}, DPR~\cite{karpukhin2020dense}, CoCondenser~\cite{gao2022unsupervised}, and ANCE~\cite{xiong2021approximate}. 

Then we use different retrievers as backbone models and optimize them using different training strategies. Three encoders as backbone retrievers to examine the generalization ability of our \method{}, including vanilla BERT~\cite{devlin2019bert}, Contriever~\cite{izacard2021unsupervised}, and AnchorDR~\cite{xie2023unsupervised}. Contriever pretrains PLMs on unlabeled text pairs by encouraging semantically similar sentences to have closer representations in the embedding space. In contrast, AnchorDR leverages the relationships between anchor texts and their linked documents to enhance pretraining. Each retriever is evaluated under three training strategies: (1) \textbf{Raw}: directly encoding both queries and documents without fine-tuning; (2) \textbf{FT}: standard supervised fine-tuning using query-document triples; and (3) \textbf{ExpandR}: it integrates LLM-based query expansion to augment dense retriever and jointly optimizes both LLM and retriever.

\textbf{Implementation Details.} 
For our query expansion model, we deploy the Meta-LLaMA-3-8B-Instruct~\cite{llama3modelcard} as the backbone. The batch size is set to 16, and the learning rate is set to $2e-5$. Optimization is performed using the AdamW optimizer. We employ LoRA~\cite{hu2022lora} to efficiently fine-tune the model for 2 epochs. The temperature for the construction of the DPO data varies across $\tau \in \{0.8, 0.9, 1.0, 1.1\}$, with each setting sampled eight times. For the dense retrievers, we utilize three retrievers with different structures: BERT~\cite{devlin2019bert}, Contriever~\cite{izacard2021unsupervised} and AnchorDR~\cite{xie2023unsupervised} as the backbone. During training, we set the batch size to 1,024 and the learning rate to $1e-5$, with the model trained for 3 epochs.

%% file: Table/dataset.tex
\begin{table}
\centering
\resizebox{\linewidth}{!}{
\begin{tabular}{l l r r r}
\toprule  
\multirow{2}{*}{\textbf{Dataset}} & \multirow{2}{*}{\textbf{Setting}} & \multicolumn{3}{c}{\textbf{\#Query}} \\
\cmidrule(lr){3-5} 
&  & {Train} & {Dev} & {Test} \\ \midrule
\multirow{2}{*}{\textbf{E5}} & LLM       & 27,000  & 3,000  & -  \\
                             &Retrieval & 637,866 & 70,874 & -  \\
\midrule

{\textbf{MS MARCO}}         & Retrieval          & -       &  - & 6,980     \\
\midrule
{\textbf{Beir}}             & Retrieval          & -       & -      & 46,379    \\
\bottomrule

\end{tabular}}
\caption{\label{tab:dataset}Statistics of the datasets used in our experiments. The E5 dataset is used for joint training of the LLM and the retriever, while MS MARCO and BEIR are used exclusively for evaluation.}
\end{table}

%% file: section/5_result.tex
\input{Table/overall_new}

\section{Evaluation Results}

This section presents the overall performance of \method{}, followed by ablation studies. Then we analyze the semantic distribution of query-document embeddings under different training strategies and evaluate the effectiveness of various reward models. A case study is provided in Appendix~\ref{app:case_study}.

\subsection{Overall Performance}

The retrieval performance measured by nDCG@10 across various baselines and training configurations is summarized in Table~\ref{tab:overall-expandr}. Additional comparisons with mainstream retriever baselines and extended evaluation results are provided in Appendix~\ref{app:more_baselines} and Appendix~\ref{app:recall@100}, respectively.

As shown in the evaluation results, \method{} achieves more than 9\% improvements over previous retriever models, such as BM25 and ANCE, highlighting its effectiveness.
By substituting different retrieval backbone models, \method{} further demonstrates strong generalization ability, consistently outperforming both zero-shot retrieval (Raw) and standard supervised fine-tuning (FT). Specifically, it achieves an average improvement of 15.6\% over Raw and 5.8\% over FT across three backbone retrievers on all tasks, validating the benefit of incorporating LLM guidance into dense retrieval.

Notably, \method{} achieves the best performance on 7 out of 15 tasks when using Contriever, and on 6 tasks with AnchorDR, indicating that its effectiveness holds even with stronger backbone retrievers. The performance gains are particularly pronounced on challenging datasets such as NQ, HotpotQA, and TREC-COVID, where bridging the semantic gap between queries and documents is more difficult. These results illustrate the capability of \method{} to mitigate the semantic mismatch in complex retrieval scenarios.
Additional results using different LLMs as the backbone for query expansion are provided in Appendix~\ref{app:qwen}, showing consistent improvements and further validating the robustness of \method{} across model variants.


\subsection{Ablation Study}\label{sec:ablation}
\input{Table/ablation_new}

\input{Figure/tsne}\label{sec:tsne_analysic}
In this subsection, we conduct comprehensive ablation studies under both Contriever and AnchorDR as backbone retrievers to understand the contribution of each component in \method{}. We evaluate the impact of different reward modeling methods, LLM optimization strategies, and retriever training.

As shown in Table~\ref{tab:ablation}, we first include two baselines: ``Query'' uses raw queries without training, and ``w/ Retriever Training'' applies contrastive training using raw queries. These settings serve as control groups to isolate the contributions of our LLM optimization and expansion-based retriever training.  In both Contriever and AnchorDR backbones, we observe substantial improvements of \method{} over these baselines, demonstrating that our joint optimization strategy yields significant gains over standard query-only training.

We further assess the role of LLM optimization by removing the DPO training. This results in a 2.73\% and 4.23\% performance drop on Contriever and AnchorDR, respectively, underscoring the importance of aligning LLM outputs with ranking preferences via preference modeling.
Additionally, removing retriever training while retaining LLM optimization significantly impairs performance (2.65 and 5.56 point drops), demonstrating that expansions optimization alone is insufficient unless the retriever is also jointly adapted to leverage them. These findings validate the core motivation of \method{} that joint optimization of generation and retrieval is key to improving retrieval performance.

Finally, we conduct ablation studies by individually removing the self-reward and retriever reward to assess the impact of each reward modeling strategy during LLM training. We observe performance degradation in both cases, especially on QA benchmarks such as NQ and HotpotQA, demonstrating their complementary benefits in enhancing generation quality and aligning with retriever preferences. Notably, removing the retriever reward results in a slightly larger drop, indicating that retrieval-guided feedback plays a more crucial role in guiding effective query expansion.


\subsection{Visualization of Alignment in the Semantic Embedding Space}
We visualize the embeddings of queries and documents using T-SNE to investigate how different query expansion strategies and retriever configurations affect their semantic alignment. Specifically, we randomly sample 10 query-document pairs and project their embeddings into a two-dimensional space. Each pair is assigned a unique color, with the query represented by a star and the document by a circle, facilitating a direct visual assessment of semantic proximity under various settings. Throughout this analysis, we employ AnchorDR as the base dense retriever to encode queries and documents.

As shown in Figure~\ref{fig:embedding_visullization}, when using original queries with the base retriever (Figure~\ref{fig:original_query}), we observe that query and document embeddings are widely scattered, suggesting a substantial semantic gap between the raw query formulation and its target document. Incorporating expansions generated by a vanilla LLM leads to modest improvements (Figure~\ref{fig:q2d_contriever}), as some queries shift closer to their corresponding documents. However, the alignment remains inconsistent, and many query-document pairs still appear poorly matched. Fine-tuning the retriever alone results in further improvement (Figure~\ref{fig:q2d_trained}), making the embedding space more compact and pulling many expanded queries closer to their paired documents. Nevertheless, the most significant alignment gain is observed when both the query expansion model and the retriever are jointly optimized via preference alignment (Figure~\ref{fig:dpo_q2d_trained}). In this setting, query-document pairs exhibit significantly tighter and more coherent clustering, suggesting that the combined optimization of the expansion model and the retriever substantially improves semantic consistency and retrieval accuracy. These observations further underscore the importance of jointly aligning both components in dense retrieval systems.

\subsection{Effectiveness of Reward Modeling in Optimizing \method{}}\label{sec:reward-analysis}
Figure~\ref{fig:characteristics} presents an evaluation of the reward model designed in \method{}, measured by the text similarity between query expansions and either LLM-generated answers or golden documents. We compare three variants: the full model (\method{}), w/o Retriever Reward, and w/o Self-Reward.

We first assess the similarity between query expansions and LLM-generated answers (Figure~\ref{fig:characteristics:ans}). \method{} w/o Retriever Reward produces expansions most aligned with LLM-generated answers, yielding the highest BLEU score. In contrast, \method{} w/o Self-Reward achieves the lowest score, indicating that relying solely on the retriever reward is less effective in guiding \method{} to align with the information in answers, which is particularly important for QA tasks. When the self-reward is incorporated, the BLEU score improves notably, demonstrating its effectiveness in enhancing the factual precision of the expansions.

We then evaluate the similarity between query expansions and ground-truth documents (Figure~\ref{fig:characteristics:posi}). \method{} w/o Retriever Reward again performs worst, suggesting that the self-reward alone is insufficient to ensure alignment with golden documents. Conversely, \method{} w/o Self-Reward performs better, showing the utility of the retriever reward in guiding the model to produce semantically relevant expansions. The full model, integrating both rewards, achieves the highest BLEU score, highlighting the complementary strengths of self-reward and retriever reward in optimizing LLMs to generate high-quality expansions.
\input{Figure/performance_rm}

%% file: Table/overall_new.tex
\begin{table*}[t!]
\centering
\resizebox{\textwidth}{!}{%
\begin{tabular}{l|cccc|ccc|ccc|ccc}
\toprule
\multirow{2}{*}{\textbf{Task}} 
& \multirow{2}{*}{\textbf{BM25}}
& \multirow{2}{*}{\textbf{DPR}}
& \multirow{2}{*}{\textbf{CoCondenser}}
& \multirow{2}{*}{\textbf{ANCE}}
& \multicolumn{3}{c|}{\textbf{BERT}} 
& \multicolumn{3}{c|}{\textbf{Contriever}} 
& \multicolumn{3}{c}{\textbf{AnchorDR}} \\
\cmidrule(lr){6-8} \cmidrule(lr){9-11} \cmidrule(lr){12-14}
& & & & 
& Raw\rlap{$\text{}^{\dagger}$} & FT\rlap{$\text{}^{\diamond}$} & ExpandR 
& Raw\rlap{$\text{}^{\dagger}$} & FT\rlap{$\text{}^{\diamond}$} & ExpandR 
& Raw\rlap{$\text{}^{\dagger}$} & FT\rlap{$\text{}^{\diamond}$} & ExpandR \\
\midrule
MS MARCO        & 22.8 & 17.7 & 16.2 & \uline{37.0} & 0.29 & 22.68\rlap{$\text{}^{\dagger}$} & 23.54\rlap{$\text{}^{\dagger}$} & 20.55 & 32.96\rlap{$\text{}^{\dagger}$} & 33.65\rlap{$\text{}^{\dagger}$} & 25.66 & 36.35\rlap{$\text{}^{\dagger}$} & \textbf{37.14}\rlap{$\text{}^{\dagger}$} \\
Trec-COVID      & \uline{65.6} & 33.2 & 40.4 & 62.1 & 3.73 & 19.72\rlap{$\text{}^{\dagger}$} & 19.12\rlap{$\text{}^{\dagger}$} & 27.45 & 30.03\rlap{$\text{}^{\dagger}$} & 47.98\rlap{$\text{}^{\dagger \diamond}$} & 51.44 & 53.71\rlap{$\text{}^{\dagger}$} & \textbf{78.85}\rlap{$\text{}^{\dagger \diamond}$} \\
NFCorpus        & \uline{32.5} & 18.9 & 28.9 & 23.4 & 2.60 & 21.02\rlap{$\text{}^{\dagger}$} & 23.98\rlap{$\text{}^{\dagger \diamond}$} & 31.73 & 32.33 & \textbf{34.80}\rlap{$\text{}^{\dagger \diamond}$} & 31.23 & 31.04 & 32.13\rlap{$\text{}^{\diamond}$} \\
NQ              & 32.9 & 47.4 & 17.8 & 42.9 & 0.40 & 15.61\rlap{$\text{}^{\dagger}$} & 29.64\rlap{$\text{}^{\dagger \diamond}$} & 25.37 & 33.72\rlap{$\text{}^{\dagger}$} & \uline{50.39}\rlap{$\text{}^{\dagger \diamond}$} & 26.24 & 40.30\rlap{$\text{}^{\dagger}$} & \textbf{55.91}\rlap{$\text{}^{\dagger \diamond}$} \\
HotpotQA        & 60.3 & 39.1 & 34.0 & 47.1 & 0.77 & 16.10\rlap{$\text{}^{\dagger}$} & 29.70\rlap{$\text{}^{\dagger \diamond}$} & 48.07 & 58.78\rlap{$\text{}^{\dagger}$} & \textbf{70.50}\rlap{$\text{}^{\dagger \diamond}$} & 52.46\rlap{$\text{}^{\diamond}$} & 47.84 & \uline{63.40}\rlap{$\text{}^{\dagger \diamond}$} \\
FiQA            & 23.6 & 11.2 & 25.1 & 29.3 & 0.59 & 11.16\rlap{$\text{}^{\dagger}$} & 15.40\rlap{$\text{}^{\dagger \diamond}$} & 24.50 & 26.06\rlap{$\text{}^{\dagger}$} & \uline{32.40}\rlap{$\text{}^{\dagger \diamond}$} & 24.04 & 28.20\rlap{$\text{}^{\dagger}$} & \textbf{34.17}\rlap{$\text{}^{\dagger \diamond}$} \\
ArguAna         & 31.5 & 17.5 & 44.4 & 40.2 & 8.19 & 39.36\rlap{$\text{}^{\dagger}$} & 37.57\rlap{$\text{}^{\dagger}$} & 37.90 & \uline{53.48}\rlap{$\text{}^{\dagger}$} & \textbf{55.39}\rlap{$\text{}^{\dagger \diamond}$} & 29.50 & 48.51\rlap{$\text{}^{\dagger}$} & 49.16\rlap{$\text{}^{\dagger}$} \\
Touche-2020     & \textbf{36.7} & 13.1 & 11.7 & 23.6 & 0.39 & 2.82\rlap{$\text{}^{\dagger}$}  & 5.89\rlap{$\text{}^{\dagger \diamond}$}  & 16.68\rlap{$\text{}^{\diamond}$} & 10.46 & 17.38\rlap{$\text{}^{\diamond}$} & 12.37 & 13.76\rlap{$\text{}^{\dagger}$} & \uline{24.53}\rlap{$\text{}^{\dagger \diamond}$} \\
CQADupStack     & 29.9 & 15.3 & 30.9 & 28.8 & 1.10 & 17.10\rlap{$\text{}^{\dagger}$} & 16.47\rlap{$\text{}^{\dagger}$} & 28.43 & 31.60\rlap{$\text{}^{\dagger}$} & 33.00\rlap{$\text{}^{\dagger \diamond}$} & 30.30 & \uline{34.72}\rlap{$\text{}^{\dagger}$} & \textbf{35.18}\rlap{$\text{}^{\dagger}$} \\
Quora           & 78.9 & 24.8 & 82.1 & 84.7 & 36.29& 77.38\rlap{$\text{}^{\dagger}$} & 72.04\rlap{$\text{}^{\dagger}$} & 83.50 & \uline{84.98}\rlap{$\text{}^{\dagger}$} & 84.67\rlap{$\text{}^{\dagger}$} & 83.49 & \textbf{85.06}\rlap{$\text{}^{\dagger}$} & 79.34 \\
DBPedia         & 31.3 & 26.3 & 21.5 & 26.5 & 1.57 & 14.08\rlap{$\text{}^{\dagger}$} & 23.05\rlap{$\text{}^{\dagger \diamond}$} & 29.16 & 36.46\rlap{$\text{}^{\dagger}$} & \textbf{42.32}\rlap{$\text{}^{\dagger \diamond}$} & 33.58 & 34.55 & \uline{40.73}\rlap{$\text{}^{\dagger \diamond}$} \\
Scidocs         & 15.8 & 7.7  & 13.6 & 11.3 & 0.70 & 6.04\rlap{$\text{}^{\dagger}$}  & 9.43\rlap{$\text{}^{\dagger \diamond}$}  & 14.91 & 14.94 & \textbf{17.85}\rlap{$\text{}^{\dagger \diamond}$} & 16.57 & 15.77 & \uline{16.82}\rlap{$\text{}^{\diamond}$} \\
FEVER           & 75.3 & 56.2 & 61.5 & 68.1 & 0.24 & 36.59\rlap{$\text{}^{\dagger}$} & 57.49\rlap{$\text{}^{\dagger \diamond}$} & 68.20 & 82.49\rlap{$\text{}^{\dagger}$} & \textbf{87.07}\rlap{$\text{}^{\dagger \diamond}$} & 62.98 & 77.43\rlap{$\text{}^{\dagger}$} & \uline{84.57}\rlap{$\text{}^{\dagger \diamond}$} \\
Climate-FEVER   & 21.4 & 14.8 & 16.9 & 19.8 & 0.61 & 11.52\rlap{$\text{}^{\dagger}$} & 24.63\rlap{$\text{}^{\dagger \diamond}$} & 15.50 & 23.04\rlap{$\text{}^{\dagger}$} & \uline{29.77}\rlap{$\text{}^{\dagger \diamond}$} & 23.44 & 26.63\rlap{$\text{}^{\dagger}$} & \textbf{31.76}\rlap{$\text{}^{\dagger \diamond}$} \\
Scifact         & 66.5 & 31.8 & 56.1 & 50.2 & 2.81 & 42.35\rlap{$\text{}^{\dagger}$} & 46.27\rlap{$\text{}^{\dagger \diamond}$} & 64.92 & \uline{68.84}\rlap{$\text{}^{\dagger}$} & \textbf{69.68}\rlap{$\text{}^{\dagger}$} & 59.84 & 60.51 & 63.43\rlap{$\text{}^{\dagger \diamond}$} \\
\midrule
Avg.BEIR14       & 43.0 & 25.5 & 34.6 & 39.9 & 4.29 & 23.63 & 29.33 & 36.88 & 41.94 & \uline{48.09} & 38.39 & 42.72 & \textbf{49.28} \\
Avg.All          & 41.7 & 25.0 & 33.4 & 39.7 & 4.02 & 23.57 & 28.95 & 35.79 & 41.34 & \uline{47.12} & 37.54 & 42.29 & \textbf{48.47} \\
\textbf{Best on} & 1 & 0 & 0 & 0 & 0 & 0 & 0 & 0 & 0 & \textbf{7} & 0 & 1 & \uline{6} \\
\bottomrule
\end{tabular}}
\caption{Overall Performance of \method{}. We follow previous work~\cite{izacard2021unsupervised} and report the average performance on 14 BEIR tasks (BEIR14) and all tasks (All). \textbf{Bold} and \uline{underlined} scores indicate the best and second-best results. $\dagger$, $\diamond$ denote significant improvements over the Raw and FT training settings of each retriever.}
\label{tab:overall-expandr}
\end{table*}

%% file: Table/ablation_new.tex
\begin{table*}[t!]
  \centering
  \resizebox{\textwidth}{!}{
  \begin{tabular}{l|cccccccc|c} 
    \toprule 
     \textbf{Model} & \textbf{MARCO} & \textbf{Trec-COVID} & \textbf{NQ} & \textbf{HotpotQA} & \textbf{FiQA} & \textbf{DBPedia} & \textbf{FEVER} & \textbf{Scifact} & \textbf{Avg.} \\ 
     \midrule
      \rowcolor{gray!10}\multicolumn{10}{c}{\textit{Contriever}} \\
      Query           & 20.55          & 27.45          & 25.37    & 48.07    & 24.50    & 29.16    & 68.20    & 64.92    & 38.53     \\
      w/ Retriever Training  & 32.96           & 30.03          & 33.72          & 58.78          & 26.06          & 36.46          & 82.49          & 68.84    & 46.17     \\ 
    \midrule
    
    \textbf{\method{}}  & \textbf{33.65} & 47.98  & \textbf{50.39} & \textbf{70.50} & \textbf{32.40}          & \textbf{42.32}          & 87.07 & 69.68 & \textbf{54.25} \\
     w/o LLM Training    & 33.45          & 38.64          & 47.20          & 66.45          & 29.74          & 40.97          & 85.18          & 70.55    & 51.52     \\ 
    w/o Retriever Training    & 25.20          & \textbf{59.66}          & 43.26 & 65.82 & 30.12 & 38.20 & 82.80          & 67.74 & 51.60 \\
    w/o Self-Reward    & 33.05          & 44.07          & 47.74         & 69.62          & 30.74          & 42.24          & \textbf{87.63} & \textbf{70.65} & 53.21 \\
    w/o Retriever Reward  & 33.47       & 42.17          & 49.75          & 69.12          & 32.12          & 40.31          & 86.52          & 69.96 & 52.92     \\
    
      \midrule
    \rowcolor{gray!10}\multicolumn{10}{c}{\textit{AnchorDR}} \\
      Query           & 25.66          & 51.44          & 26.24    & 52.46    & 24.04    & 33.58    & 62.98    & 59.84    & 42.03     \\
      w/ Retriever Training  & 36.35           & 53.71          & 40.30          & 47.84          & 28.20          & 34.55          & 77.43          & 60.51    & 47.36     \\ 
    \midrule

    \textbf{\method{}}  & \textbf{37.14} & \textbf{78.85}  & \textbf{55.91} & \textbf{63.40} & \textbf{34.17}   & \textbf{40.73}   & \textbf{84.57} & \textbf{63.43} & \textbf{57.28} \\
    w/o LLM Training                   & 35.17   & 70.56   & 51.24   & 59.22   & 29.84   & 36.11   & 80.69   & 61.58   & 53.05   \\
    w/o Retriever Training    & 29.59   & 78.50   & 42.30   & 57.41   & 24.91   & 38.67   & 79.00   & 63.40   & 51.72   \\
    w/o Self-Reward           & 36.56   & 75.75   & 54.81   & 62.74   & 32.31   & 40.42   & 84.41   & 63.07   & 56.25   \\
    w/o Retriever Reward      & 37.07   & 73.75   & 55.19   & 61.59   & 32.97   & 40.20   & 82.02   & 62.47   & 55.65   \\

      \bottomrule
    \end{tabular}}
  \caption{Ablation Analysis of Key Components in \method{} on Contriever and AnchorDR.
We examine the contributions of LLM training, retriever training, and reward modeling to retrieval performance on 8 important datasets in BEIR. MARCO denotes the MS MARCO dataset.}
  \label{tab:ablation}
\end{table*}

%% file: Figure/tsne.tex
\begin{figure}[t]
    \centering
        \subfigure[Raw Query.]{
        \label{fig:original_query}
        \includegraphics[width=0.46\linewidth]{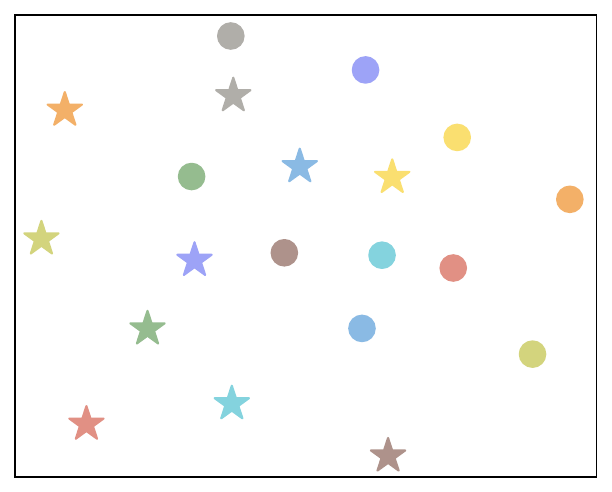}
    }
        \subfigure[Vanilla LLM.]{
        \label{fig:q2d_contriever}
        \includegraphics[width=0.46\linewidth]{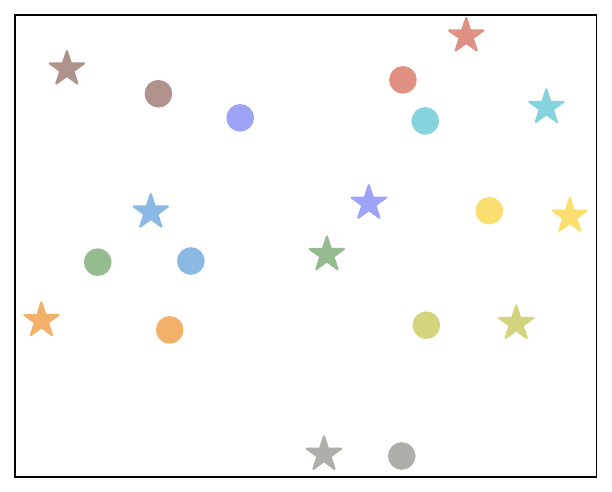}
    }
    \subfigure[\method{} (w/o DPO).]{
        \label{fig:q2d_trained}
        \includegraphics[width=0.46\linewidth]{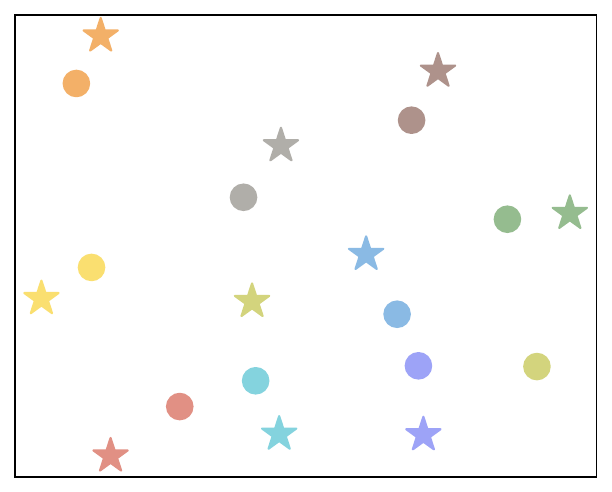}
    }
    \subfigure[\method{}.]{
        \label{fig:dpo_q2d_trained}
        \includegraphics[width=0.46\linewidth]{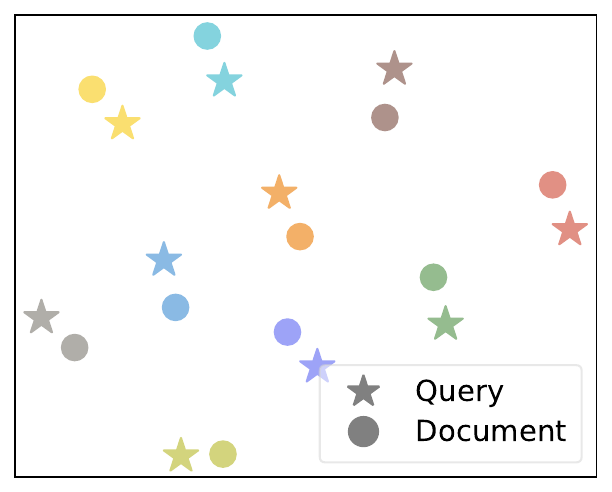}
    }
    
    \caption{Embedding Visualization of Different Models.}
    \label{fig:embedding_visullization}
\end{figure}


%% file: Figure/performance_rm.tex
\begin{figure}[t]
    \centering
        \subfigure[Similarity with Answers.]{
        \label{fig:characteristics:ans}
        \includegraphics[width=0.45\linewidth,height=3.35cm]{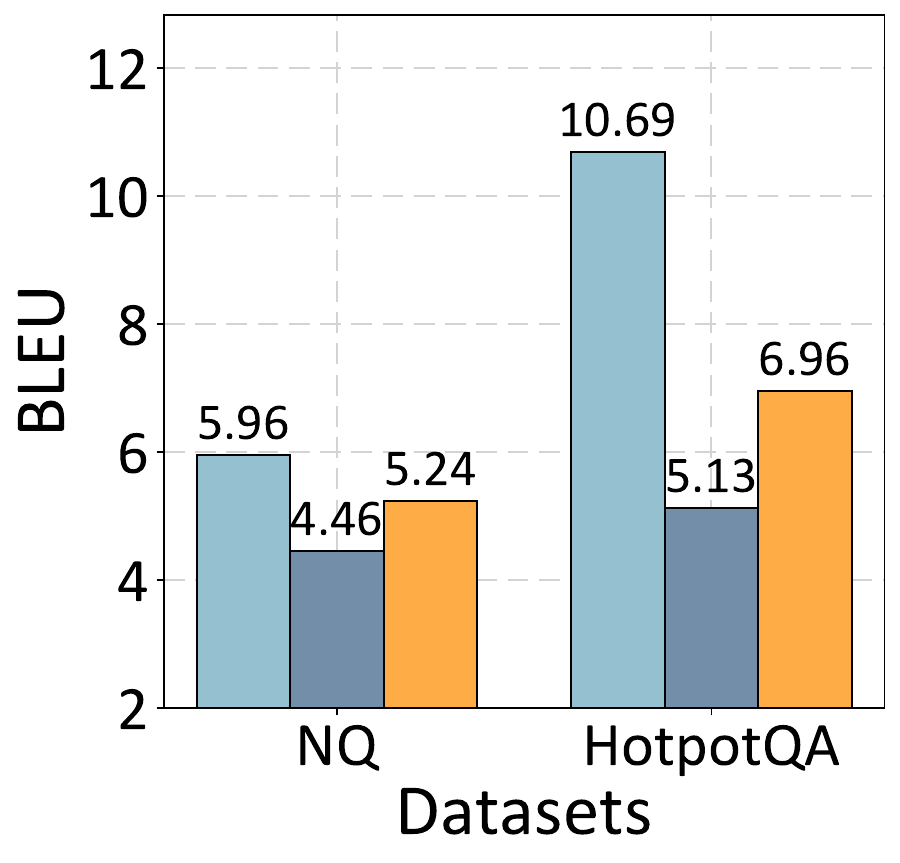}
    }
    \subfigure[Similarity with Golden Documents.]{
        \label{fig:characteristics:posi}
        \includegraphics[width=0.45\linewidth]{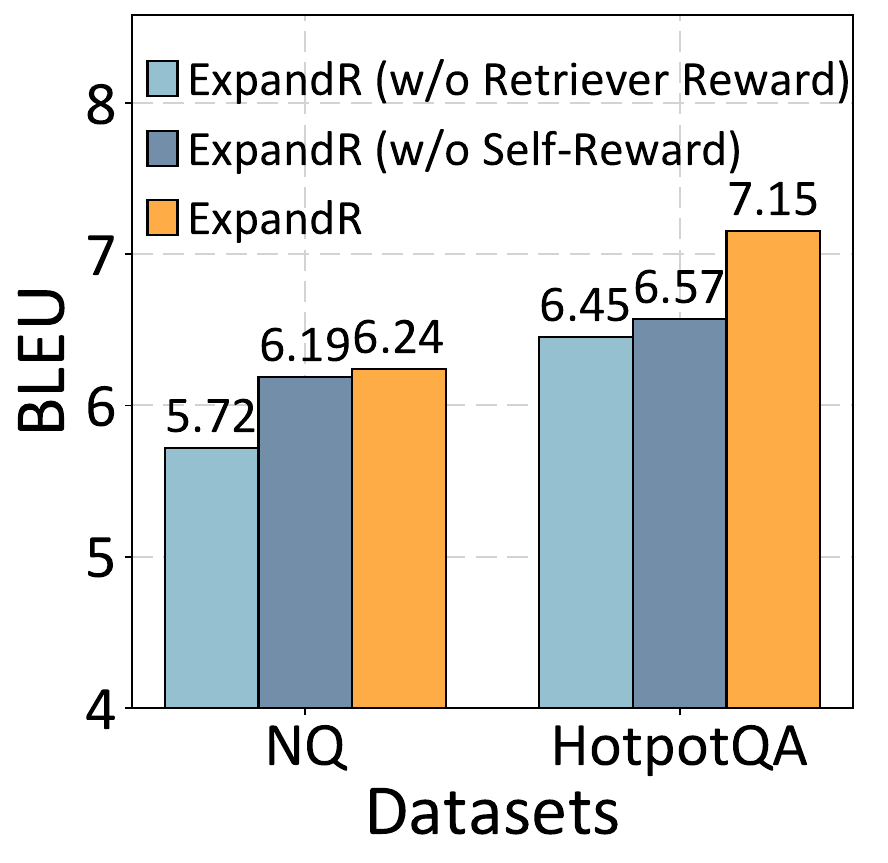}
    }
    \caption{Effect of Reward Modeling on the Semantic Alignment of Query Expansions.}
    \label{fig:characteristics}
\end{figure}

%% file: section/6_conclusion.tex
\section{Conclusion}

This paper presents \method{}, a joint optimization framework that leverages LLM-guided query expansions to enhance retriever training. By jointly training dense retrievers and LLMs, \method{} improves the effectiveness and compatibility of query expansions within retrieval systems. Experimental results demonstrate that \method{} consistently boosts performance and offers a new perspective on end-to-end alignment between generative and retrieval components in retrieval pipelines.

%% file: section/7_limitation.tex
\section*{Limitations}

Despite the effectiveness of \method{} in improving dense retrieval through LLM-guided query expansions, several limitations remain. First, the quality of expansions is still constrained by the generative capacity of the LLM. If the LLM produces low-quality or biased expansions, the downstream retriever may be misled, even with reward-based supervision. Additionally, although the end-to-end optimization improves alignment between generation and retrieval, it introduces additional computational overhead from both expansion generation and joint training.

%% file: section/8_appendix.tex
\section{Appendix}\label{appendix}
\subsection{License}
The authors of 4 out of the 15 datasets in the BEIR benchmark (NFCorpus, FiQA-2018, Quora, Climate-Fever) and the authors of ELI5 in the E5 dataset do not report the dataset license in the paper or a repository. We summarize the licenses of the remaining datasets as follows.

MS MARCO (MIT License); FEVER, NQ, and DBPedia (CC BY-SA 3.0 license); ArguAna and Touché-2020 (CC BY 4.0 license); CQADupStack and TriviaQA (Apache License 2.0); SciFact (CC BY-NC 2.0 license); SCIDOCS (GNU General Public License v3.0); HotpotQA and SQuAD (CC BY-SA 4.0 license); TREC-COVID (Dataset License Agreement).

All these licenses and agreements permit the use of their data for academic purposes.

\subsection{Additional Experimental Details}\label{app:experiment_detail}
This subsection outlines the components of the training data and presents the prompt templates used in the experiments.

\textbf{Training Datasets.} Following the setup of \citet{wang2024improving}, we use the following datasets: ELI5 (sample ratio 0.1)~\cite{fan2019eli5}, HotpotQA~\cite{yang2018hotpotqa}, FEVER~\cite{thorne2018fever}, MS MARCO passage ranking (sample ratio 0.5) and document ranking (sample ratio 0.2)~\cite{bajaj2016ms}, NQ~\cite{karpukhin2020dense}, SQuAD~\cite{karpukhin2020dense}, and TriviaQA~\cite{karpukhin2020dense}. In total, we use 808,740 training examples.

\textbf{Prompt Templates.} Table~\ref{tab:prompt_template} lists all the prompts used in this paper. In each prompt, ``query'' refers to the input query for which query expansions are generated, while ``Related Document'' denotes the ground truth document relevant to the original query. We observe that, in general, the model tends to generate introductory phrases such as ``Here is a passage to answer the question:'' or ``This is the answer to the query:''. Before using the model outputs as query expansions or answer signals, we first filter out these introductory phrases to ensure cleaner and more precise expansion results.
\input{Table/prompt}

\input{Table/more_retriever}

\subsection{Comparison with Mainstream Retrievers}\label{app:more_baselines}

To further contextualize the performance of \method{}, we compare it with a range of widely used dense retrievers on the BEIR and MS MARCO datasets, as shown in Table~\ref{tab:retriever-comparison}. The baselines include RocketQA~\cite{ren2021rocketqav2}, BGE-M3-EN~\cite{chen2024bge}, TAS-B~\cite{hofstatter2021efficiently}, Gen-Q, ColBERT~\cite{khattab2021relevance}, E5~\cite{wang2022text}, WebDRO~\cite{han2023enhancing}, and Nomic-Embed~\cite{nussbaum2024nomic}, covering both general-purpose and specialized retrieval models.The base retriever of the \method{} method is AnchorDR.

\method{} achieves the highest average performance across all datasets (48.5\%), consistently outperforming all baselines. Even when excluding MS MARCO—which some retrievers may be specifically optimized for—\method{} retains its leading position with an average score of 49.3\%, suggesting strong generalization across a wide range of domains and task formats.

Among the baselines, E5 and Nomic-Embed stand out as strong retrievers. E5 performs competitively on several QA-style datasets such as MS MARCO and NQ, while Nomic-Embed excels on tasks like ArguAna and HotpotQA. However, both models exhibit noticeable performance drops on other benchmarks—for instance, Nomic-Embed underperforms on MS MARCO and Touche-2020—indicating limitations in generalization. In contrast, \method{} demonstrates more consistent performance across the board, achieving top-tier results without compromising on robustness. This highlights the robustness and generalizability of our approach across diverse retrieval scenarios.

\input{Table/overall_recall}

\subsection{Evaluating Retrieval Completeness through Recall@100}\label{app:recall@100}

To more comprehensively assess the retrieval capabilities of \method{}, we report its performance under the Recall@100 metric on both the BEIR and MS MARCO datasets. This metric reflects the model's ability to retrieve a broad set of relevant documents, complementing earlier evaluations based on ranking accuracy. The results are presented in Table~\ref{tab:overall_recall}.

Across all retriever backbones, \method{} consistently achieves the highest Recall@100 scores, surpassing both the original query (Raw) and supervised retriever (FT) baselines. The improvements are particularly notable on complex multi-hop and fact-seeking datasets such as NQ, HotpotQA, and FEVER, where purely lexical signals are often insufficient for comprehensive retrieval.

These findings suggest that \method{} not only improves ranking precision but also significantly enhances semantic recall, demonstrating its ability to uncover a wider range of relevant documents. This further validates the robustness and general applicability of our LLM-augmented strategy across diverse retrieval scenarios.

\subsection{Robustness under Different LLM Backbones}\label{app:qwen}
\input{Table/qwen_overall}

To examine the robustness of \method{} across different language model backbones, we replace the LLM used for query expansion with Qwen2.5-7B-Instruct~\cite{yang2024qwen2}, a high-quality Chinese-English bilingual model trained with instruction tuning. We keep Contriever as the base retriever. The results are shown in Table~\ref{tab:qwen-extend}.

The results show that \method{} consistently outperforms both the original query baseline (Raw) and the supervised retriever trained with raw queries (FT), achieving the best performance on 14 out of 15 datasets. The performance trend closely mirrors that observed in our original experiments using LLaMA, indicating that the improvements are not tied to a specific LLM architecture. Instead, \method{} captures a generally effective joint optimization strategy that transfers well across different language models.

\input{Figure/improvement}
\subsection{Query Expansion Quality of \method{}}\label{app:analysis}
This section evaluates the quality of query expansion of \method{}. As shown in Figure~\ref{fig:imp}, we randomly select 100 samples from each dataset to assess the improvement in retrieval performance before and after applying \method{}.

Overall, the evaluation results demonstrate that \method{} consistently improves retrieval performance in both unsupervised (Figure~\ref{fig:imp:unsupervised}) and supervised (Figure~\ref{fig:imp:supervised}) settings. However, for the MS MARCO dataset, \method{} demonstrates limited effectiveness in the supervised setting. This can be attributed to the fact that MS MARCO provides higher-quality training signals, allowing the dense retriever to learn sufficient matching signals from relevance labels. In contrast, \method{} leads to more substantial performance improvements on the NQ and HotpotQA datasets. This indicates that \method{} provides essential matching signals for dense retrievers, particularly in retrieval scenarios where high-quality training signals are scarce.

\input{Table/cross_retriever}

\subsection{Generalization Analysis of Ranking-Aligned LLM Expansions}
To examine the generalizability of our ranking-aligned query expansions beyond the retriever used during training, we evaluate \method{} under two structurally distinct dense retrievers—AnchorDR and BGE-large-1.5—while keeping the reward signals derived from Contriever fixed.

As shown in Table~\ref{tab:cross-retriever}, the results show that retrieval using expansions generated by \method{} consistently yields better performance than using either the original queries or expansions produced by a vanilla LLM, across both retrievers. Although the LLM is optimized using reward signals from Contriever, it achieves strong performance under both AnchorDR and BGE, obtaining the best results on 12 out of 15 datasets in each setting. Notably, even on BGE—an already highly effective retriever—\method{} still achieves further gains, indicating that the learned expansions do not simply overfit to the behavior of a specific model, but instead capture a transferable ranking preference that generalizes across different retrieval architectures.

\subsection{More Insights into the Self-Reward}

\input{Figure/length_analysis}

While the primary purpose of introducing the self-reward is to enhance the semantic relevance between the generated expansions and the gold answer, we observe that it also serves as an effective regularizer for controlling generation quality. Specifically, we compare the average lengths of the expansions produced by three variants of our model. As shown in Figure~\ref{fig:characteristics:length}, removing the self-reward leads to significantly longer generations, which are not necessarily more informative and may introduce hallucinated or off-topic content—a known issue in preference-based tuning methods such as DPO.

With the self-consistency signal in place, the model generates shorter and more focused expansions. To further assess the semantic faithfulness of these generations, we conduct a natural language inference (NLI) based entailment evaluation. As shown in Table~\ref{tab:nli_hallucination}, although removing the self-reward increases the average length, it results in lower entailment scores, suggesting reduced semantic alignment with the gold answer. In contrast, the full model—trained with both the retriever-based and self-rewards—achieves the highest entailment scores while keeping the generation length moderate, indicating a better balance between informativeness and faithfulness.

These results suggest that the self-reward not only enhances \(\log P(d^{\text{exp}} \mid q; \Theta)\), but also implicitly constrains the LLM from over-generating, thereby mitigating hallucination and improving the overall quality of the query expansions during DPO training.
\input{Table/nli_analyze}

\subsection{Case Study}\label{app:case_study}
To further demonstrate the effectiveness of \method{}, we conduct a case study by randomly sampling a query from the evaluation dataset. We then compare retrieval performance using the raw queries, expanded queries by vanilla LLM, and expanded queries by \method{}.

As shown in Table~\ref{tab:case_study}, query expansion significantly improves retrieval effectiveness over using the raw query, with both LLM-generated variants achieving higher nDCG@10. While the vanilla LLM introduces relevant terms such as ``temperature'' and ``humidity'', its expansions are often verbose and include redundant or inconsistent content (e.g., conflicting temperature ranges). This reflects a lack of alignment between generation and retrieval utility.

In contrast, \method{} produces expansions that are more concise and semantically aligned with the golden passage, incorporating key concepts such as ``human behavior'', ``environmental factors'', and ``virus transmission''. These expansions better match the relevance signals favored by the retriever, leading to improved ranking performance. This example illustrates how preference-guided fine-tuning in \method{} enables the LLM to generate expansions that are both informative and behaviorally aligned with the retrieval model.
\input{Table/case_study}

%% file: Table/prompt.tex
\begin{table}[t]
\small
\definecolor{titlecolor}{RGB}{218, 227, 245}
\setlength{\arrayrulewidth}{0.7pt}
\begin{tabular}{|p{0.95\columnwidth}|}
\hline  
\multicolumn{1}{|c|}{\cellcolor{titlecolor}\textbf{Query Expansion}} \\

\uline{\textbf{Prompt for Q2D:}} \\
Please write a passage to answer the question:\\
Question: \{\}\\
Passage:\\
\hline
\multicolumn{1}{|c|}{\cellcolor{titlecolor}\textbf{Question Answering}} \\
\uline{\textbf{Prompt for Q2A:}} \\
You are given a query and a related document. Based on the query, generate a direct and relevant answer using the information in the document. If the query is a statement, expand on it. If it is a question, provide a direct answer. Avoid any extra description or irrelevant content.\\
Query: \{\}\\
Related Document: \{\}\\
Answer:\\

\hline
\end{tabular}
\caption{Prompt Templates Used in \method{}. These prompts are used to generate query expansion results and produce the responses to answer the question.}
\label{tab:prompt_template}
\end{table}

%% file: Table/more_retriever.tex
\begin{table*}[ht]
\centering
\resizebox{\textwidth}{!}{
\begin{tabular}{l|cccccccc|c}
\toprule
\textbf{Task} & \textbf{RocketQA} & \textbf{BGE-M3-EN} & \textbf{TAS-B} & \textbf{Gen-Q} & \textbf{ColBERT} & \textbf{E5} & \textbf{WebDRO} & \textbf{Nomic-Embed} & \textbf{\method{}} \\
\midrule
MS MARCO      & 23.2  & 35.2  & 40.8  & 40.8  & 40.1  & \textbf{43.1}  & 40.6  & 26.4  & 37.1 \\
Trec-COVID    & 67.5  & 44.6  & 48.1  & 61.9  & 67.7  & 61.7  & 78.0  & 67.1  & \textbf{78.9} \\
NFCorpus      & 29.3  & 32.7  & 31.9  & 31.9  & 30.5  & 35.1  & 31.2  & \textbf{35.5}  & 32.1 \\
NQ            & 59.5  & 29.8  & 46.3  & 35.8  & 52.4  & \textbf{60.0}  & 47.2  & 51.2  & 55.9 \\
HotpotQA      & 35.6  & 68.3  & 58.4  & 53.4  & 59.3  & 52.4  & 57.4  & \textbf{69.1}  & 63.4 \\
FiQA          & 30.2  & 28.3  & 30.0  & 30.8  & 31.7  & 37.9  & 28.4  & \textbf{37.8}  & 34.2 \\
ArguAna       & 45.1  & \textbf{61.5}  & 42.9  & 49.3  & 23.3  & 51.4  & 48.0  & 54.2  & 49.2 \\
Touche-2020   & 24.7  & 13.5  & 16.2  & 18.2  & 20.2  & \textbf{28.3}  & 27.6  & 19.0  & 24.5 \\
CQADupStack   & 19.3  & \textbf{40.2}  & 31.4  & 34.7  & 35.0  & 28.3  & 35.2  & 49.6  & 35.2 \\
Quora         & 31.2  & \textbf{88.7}  & 83.5  & 83.0  & 85.4  & 87.9  & 85.8  & 88.4  & 79.3 \\
DBPedia       & 35.6  & 19.0  & 38.4  & 32.8  & 39.2  & 33.8  & 38.1  & 39.4  & \textbf{40.7} \\
Scidocs       & 16.5  & 9.6   & 14.9  & 14.3  & 14.5  & 19.0  & 15.3  & \textbf{19.2}  & 16.8 \\
FEVER         & 67.6  & 64.3  & 70.0  & 66.9  & 77.1  & 58.2  & 70.9  & 60.3  & \textbf{84.6} \\
C-FEVER       & 18.0  & 18.3  & 22.8  & 17.5  & 18.4  & 15.4  & 18.9  & 27.0  & \textbf{31.8} \\
Scifact       & 56.8  & 71.5  & 64.3  & 64.4  & 67.1  & \textbf{73.1}  & 62.2  & 71.8  & 63.4 \\
\midrule
Avg.BEIR14 & 38.3  & 42.2  & 42.8  & 42.5  & 44.4  & 45.9  & 46.0  & 49.2  & \textbf{49.3} \\
Av.All    & 37.3  & 41.7  & 42.7  & 42.4  & 44.1  & 45.7  & 45.6  & 47.7  & \textbf{48.5} \\
\bottomrule
\end{tabular}
}
\caption{Performance Comparison of More Mainstream Retriever Baselines on the Beir and MS MARCO Datasets (nDCG@10). The base retriever of the \method{} method is AnchorDR.}
\label{tab:retriever-comparison}
\end{table*}

%% file: Table/overall_recall.tex
\begin{table*}[t!]
\centering
\resizebox{\textwidth}{!}{%
\begin{tabular}{l|cccc|ccc|ccc|ccc}
\toprule
\multirow{2}{*}{\textbf{Task}} 
& \multirow{2}{*}{\textbf{BM25}}
& \multirow{2}{*}{\textbf{DPR}}
& \multirow{2}{*}{\textbf{CoCondenser}}
& \multirow{2}{*}{\textbf{ANCE}}
& \multicolumn{3}{c|}{\textbf{BERT}} 
& \multicolumn{3}{c|}{\textbf{Contriever}} 
& \multicolumn{3}{c}{\textbf{AnchorDR}} \\
\cmidrule(lr){6-8} \cmidrule(lr){9-11} \cmidrule(lr){12-14}
& & & & 
& Raw & FT & ExpandR 
& Raw & FT & ExpandR 
& Raw & FT & ExpandR \\
\midrule
MS MARCO      & 65.8  & 55.2   & 58.2  & 83.8  & 3.32  & 67.29 & 69.06 & 67.19 & 82.81 & 83.64 & 74.95 & \uline{84.56} & \textbf{84.83} \\
Trec-COVID    & \textbf{49.8}  & \uline{21.2}   & 7.0   & 9.6   & 0.71  & 2.05  & 3.30  & 3.68  & 3.19  & 6.58  & 10.70 & 10.67 & 14.44 \\
NFCorpus      & 25.0  & 20.8   & 29.1  & 22.3  & 8.66  & 21.40 & 25.79 & 29.41 & 15.97 & \textbf{34.07} & 28.72 & 28.93 & \uline{30.78} \\
NQ            & 76.0  & 88.0   & 67.9  & 82.2  & 2.81  & 65.82 & 83.23 & 77.12 & 88.18 & \textbf{94.88} & 80.42 & 89.67 & \uline{94.30} \\
HotpotQA      & 74.0  & 59.1   & 54.7  & 58.8  & 5.97  & 42.57 & 60.95 & 70.45 & 75.64 & \textbf{87.33} & 65.86 & 66.89 & \uline{78.72} \\
FiQA          & 53.9  & 34.2   & 60.3  & 58.2  & 4.66  & 39.45 & 47.52 & 56.19 & 61.04 & \textbf{70.03} & 54.89 & 61.07 & \uline{65.44} \\
ArguAna       & 94.2  & 75.1   & 93.0  & 92.3  & 45.73 & 95.45 & 95.38 & 90.11 & \uline{98.43} & \textbf{99.00} & 80.65 & 96.51 & 96.80 \\
Touche-2020   & \textbf{53.8}  & 30.1   & 27.1  & 45.2  & 1.33  & 14.30 & 30.37 & 37.36 & 31.52 & 46.35 & 39.91 & 38.30 & \uline{47.00} \\
CQADupStack   & 60.6  & 40.3   & 60.3  & 57.1  & 7.05  & 42.81 & 42.78 & 61.40 & 65.20 & \textbf{67.39} & 62.41 & \uline{66.44} & 66.37 \\
Quora         & 97.3  & 47.0   & 98.5  & \uline{98.6}  & 70.10 & 96.96 & 96.06 & 98.71 & \textbf{99.09} & 93.55 & 95.71 & 98.11 & 96.15 \\
DBPedia       & 39.8  & 34.9   & 34.8  & 30.8  & 3.85  & 25.92 & 34.69 & 45.29 & 48.22 & \textbf{54.00} & 43.94 & 43.73 & \uline{48.83} \\
Scidocs       & 35.6  & 21.9   & 34.1  & 25.2  & 5.67  & 22.55 & 27.28 & 35.99 & \uline{37.10} & \textbf{40.50} & 36.99 & 35.15 & 36.78 \\
FEVER         & 93.1  & 84.0   & 89.6  & 91.1  & 1.91  & 78.48 & 88.69 & 93.56 & \uline{95.93} & \textbf{96.94} & 93.65 & 93.09 & 95.05 \\
C-FEVER       & 43.6  & 39.0   & 37.0  & 45.6  & 4.23  & 44.01 & 56.84 & 44.14 & 58.56 & \uline{64.56} & 60.08 & 60.25 & \textbf{64.81} \\
Scifact       & 90.8  & 72.7   & 91.4  & 81.4  & 22.39 & 80.36 & 81.47 & 92.60 & \uline{94.00} & \textbf{96.00} & 90.77 & 91.43 & 93.43 \\
\midrule
Avg.BEIR14    & 63.4  & 48.3   & 56.1  & 57.0  & 13.22 & 48.01 & 55.31 & 59.71 & 62.29 & \textbf{67.94} & 60.34 & 62.87 & \uline{66.35} \\
Avg.All       & 63.6  & 47.7   & 56.2  & 58.8  & 12.56 & 49.29 & 56.23 & 60.21 & 63.66 & \textbf{68.99} & 61.31 & 64.32 & \uline{67.58} \\
\textbf{Best on} &   2   &  0    &  0    &  0    &  0    &    0   &   0    &  0     &  1     &   10    &  0     &   0    &  2      \\
\bottomrule
\end{tabular}}
\caption{Overall Performance of \method{} on Recall@100.}
\label{tab:overall_recall}
\end{table*}

%% file: Table/qwen_overall.tex
\begin{table}[ht]
\small
\centering
\begin{tabular}{l|ccc}
\toprule

 \multirow{2}{*}{\textbf{Task }} & \multicolumn{3}{c}{\textbf{Contriever}} \\ \cmidrule{2-4}
      & Raw & FT & \method{} \\

\midrule
MS MARCO    & 20.55 & 32.96 & \textbf{33.32} \\
Trec-COVID  & 27.45 & 30.03 & \textbf{48.18} \\
NFCorpus    & 31.73 & 32.33 & \textbf{34.58} \\
NQ          & 25.37 & 33.72 & \textbf{50.86} \\
HotpotQA    & 48.07 & 58.78 & \textbf{70.04} \\
FiQA        & 24.50 & 26.06 & \textbf{31.98} \\
ArguAna     & 37.90 & 53.48 & \textbf{55.15} \\
Touche-2020 & 16.68 & 10.46 & \textbf{18.09} \\
CQADupStack & 28.43 & 31.60 & \textbf{32.95} \\
Quora       & 83.50 & \textbf{84.98} & 84.58 \\
DBPedia     & 29.16 & 36.46 & \textbf{41.47} \\
Scidocs     & 14.91 & 14.94 & \textbf{17.48} \\
FEVER       & 68.20 & 82.49 & \textbf{87.21} \\
C-FEVER     & 15.50 & 23.04 & \textbf{30.50} \\
Scifact     & 64.92 & 68.84 & \textbf{70.00} \\
\midrule
Avg.BEIR14  & 36.88 & 41.94 & \textbf{48.08} \\
Avg.All     & 35.79 & 41.34 & \textbf{47.09} \\
\textbf{Best on} & 0 & 1 & \textbf{14} \\
\bottomrule
\end{tabular}
%
\caption{Extended Comparison Results under Qwen2.5-7B-Instruct (nDCG@10). The basic retriever in this experiment is Contriever.}
\label{tab:qwen-extend}
\end{table}

%% file: Figure/improvement.tex

\begin{figure}[t]
    \centering
    \subfigure[Unsupervised Dense Retriever.]{
        \label{fig:imp:unsupervised}
        \includegraphics[width=0.8\linewidth]{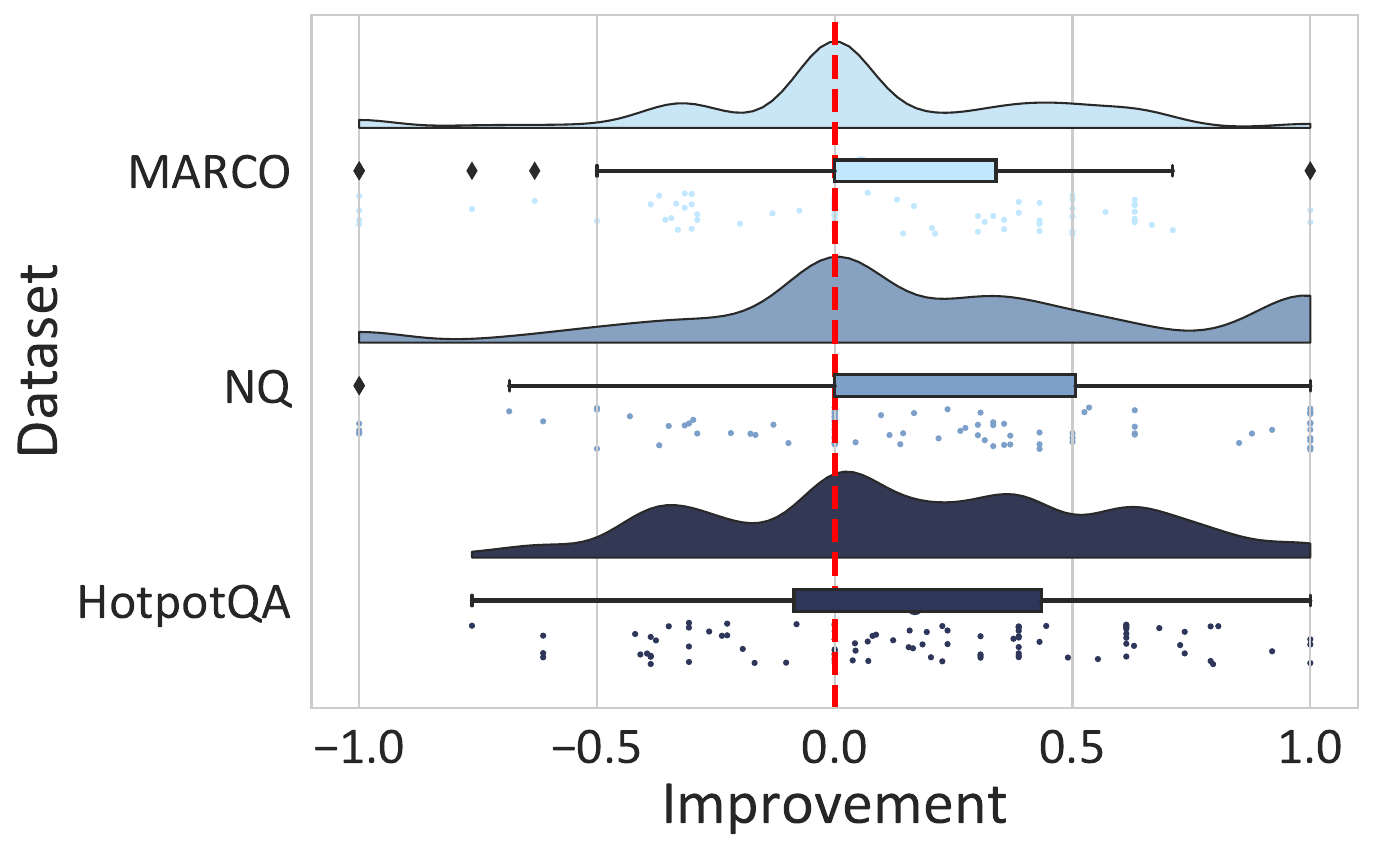}
    }
    \subfigure[Supervised Dense Retriever.]{
        \label{fig:imp:supervised}
        \includegraphics[width=0.8\linewidth]{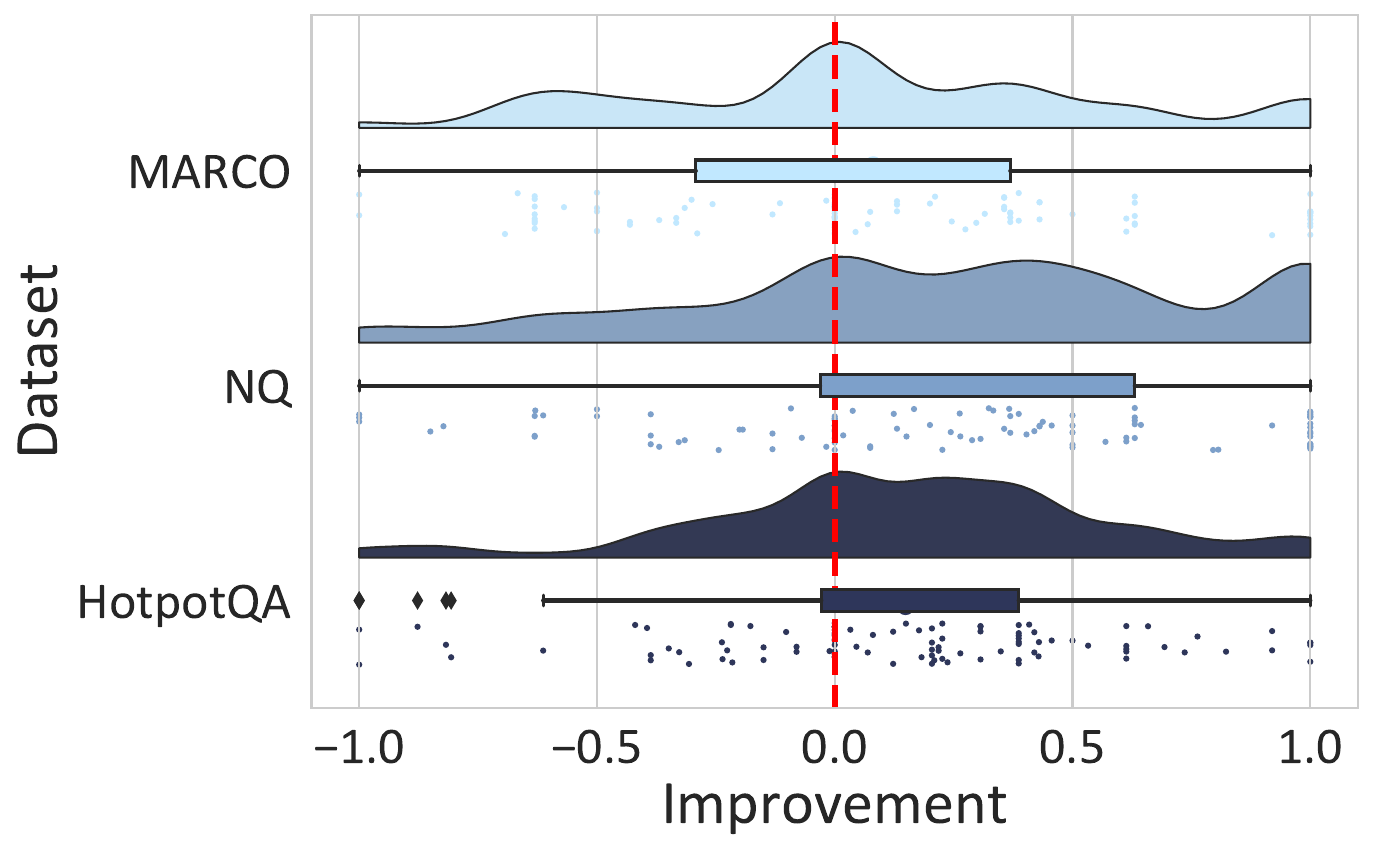}
    }
    
    \caption{Improvements of \method{} in Both Unsupervised and Supervised Dense Retrievers. We plot the change of nDCG@10 scores before and after the query expansion using our \method{} model.}
    \label{fig:imp}
\end{figure}

%% file: Table/cross_retriever.tex
\begin{table*}[ht]
\small
\centering
\begin{tabular}{l|ccc|ccc}
\toprule
\multirow{2}{*}{\textbf{Task }} & \multicolumn{3}{c|}{\textbf{AnchorDR}} & \multicolumn{3}{c}{\textbf{BGE-large-1.5}} \\
              & Query & Vanilla LLM & ExpandR & Query & Vanilla LLM & ExpandR \\
\midrule
MS MARCO      & 25.7  & 28.9  & \textbf{29.4}   & \textbf{42.0}  & 39.4  & 40.3 \\
Trec-COVID    & 51.4  & \textbf{77.9}  & 77.1   & 64.5  & 77.8  & \textbf{78.5} \\
NFCorpus      & 31.2  & 31.3  & \textbf{31.4}   & 36.8  & 37.2  & \textbf{39.3} \\
NQ            & 26.2  & 39.2  & \textbf{43.0}   & 51.7  & 59.6  & \textbf{60.8} \\
HotpotQA      & 52.5  & 58.0  & \textbf{59.3}   & 74.3  & 75.2  & \textbf{76.7} \\
FiQA          & 24.0  & 24.9  & \textbf{25.4}   & 44.3  & 44.3  & \textbf{46.2} \\
ArguAna       & \textbf{29.5}  & 28.0  & 28.2   & \textbf{63.5}  & 61.6  & 62.6 \\
Touche-2020   & 12.4  & 23.5  & \textbf{25.6}   & 24.2  & 25.3  & \textbf{26.3} \\
CQADupStack   & 30.3  & 31.1  & \textbf{31.6}   & 41.7  & 42.2  & \textbf{42.6} \\
Quora         & \textbf{83.5}  & 63.2  & 66.4   & \textbf{89.0}  & 87.9  & 88.0 \\
DBPedia       & 33.6  & 38.8  & \textbf{39.3}   & 42.1  & 45.1  & \textbf{45.2} \\
Scidocs       & 16.6  & 16.9  & \textbf{17.0}   & 20.9  & 22.9  & \textbf{23.7} \\
FEVER         & 63.0  & 77.5  & \textbf{79.7}   & 84.6  & 86.5  & \textbf{88.6} \\
C-FEVER       & 23.4  & 29.7  & \textbf{30.0}   & 28.4  & 30.6  & \textbf{31.7} \\
Scifact       & 59.8  & 62.4  & \textbf{63.2}   & 73.5  & 75.1  & \textbf{75.3} \\
\midrule
Avg.BEIR14    & 38.4  & 43.3  & \textbf{44.1}   & 52.8  & 55.1  & \textbf{56.1} \\
Avg.All       & 37.5  & 42.4  & \textbf{43.1}   & 52.1  & 54.0  & \textbf{55.1} \\
\textbf{Best on}       & 2     &  1    &  12    & 3     & 0     & 12      \\
\bottomrule
\end{tabular}
\caption{Cross-Retriever Evaluation of Ranking-Aligned Expansions (nDCG@10).}
\label{tab:cross-retriever}
\end{table*}

%% file: Figure/length_analysis.tex
\begin{figure}[t]
  \centering
  \includegraphics[width=0.75\linewidth]{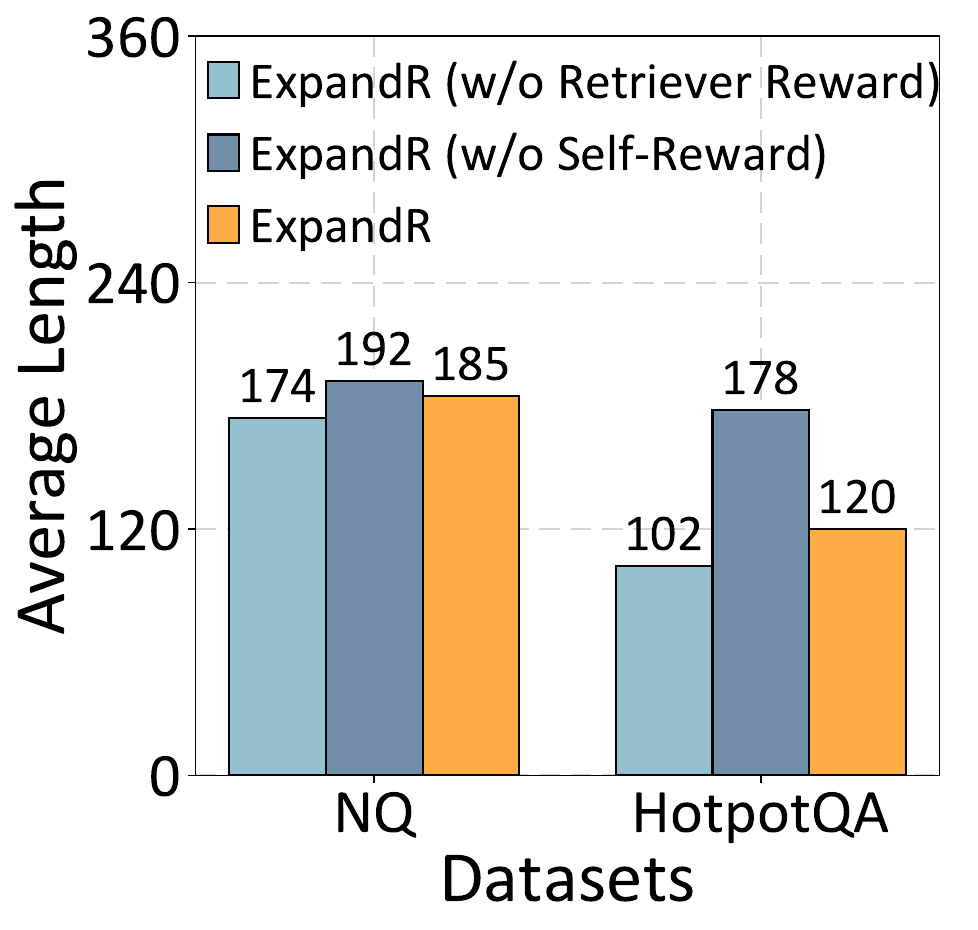}
  \caption{Average Length of Query Expansions Generated by Different Models.}
  \label{fig:characteristics:length}
\end{figure}

%% file: Table/nli_analyze.tex
\begin{table*}[ht]
\centering
\small
\begin{tabular}{l|cc|cc}
\toprule
\multirow{2}{*}{\textbf{Model}} & \multicolumn{2}{c|}{\textbf{NQ}} & \multicolumn{2}{c}{\textbf{HotpotQA}} \\
& NLI Score & Avg. Length & NLI Score & Avg. Length \\
\midrule
Vanilla LLM & 6.44 & 221.76 & 16.38 & 129.63 \\
\method{} (w/o Retriever Reward) & 8.12 & \textbf{174.20} & 17.81 & \textbf{102.60} \\
\method{} (w/o Self-Reward) & 6.65 & 192.76 & 13.75 & 178.29 \\
\method{} & \textbf{8.64} & 185.11 & \textbf{18.67} & 120.52 \\
\bottomrule
\end{tabular}
\caption{Comparison of NLI Entailment Scores and Average Lengths of Extensions Generated by Different Models.}
\label{tab:nli_hallucination}
\end{table*}

%% file: Table/case_study.tex
\begin{table*}
\centering
%
\small
\begin{tabular}{>{\raggedright\arraybackslash}p{0.96\textwidth}}
\toprule
\textbf{Query:} How does the coronavirus respond to changes in the weather? \\
\textbf{Golden Pagssage:} Epidemics ... occur during the winter months. ... Two major contributing \sethlcolor{lightblue}\hl{factors} are the \sethlcolor{lightyellow}\hl{changes} in \sethlcolor{lightblue}\hl{environmental} parameters and \sethlcolor{lightblue}\hl{human behavior}. Studies have revealed the effect of \sethlcolor{lightgreen}\hl{temperature} and \sethlcolor{lightgreen}\hl{humidity} on respiratory \sethlcolor{lightblue}\hl{virus stability} and \sethlcolor{lightblue}\hl{transmission} rates. More recent research highlights the importance of the \sethlcolor{lightblue}\hl{environmental factors}, especially \sethlcolor{lightgreen}\hl{temperature} and \sethlcolor{lightgreen}\hl{humidity} ...   \\
\midrule
\multicolumn{1}{c}{\textbf{Raw Query}} \\ 
\textbf{nDCG@10:} \textcolor{darkgray}{22.01\%} \\
\textbf{Original Query}: How does the \sethlcolor{lightyellow}\hl{coronavirus} respond to \sethlcolor{lightyellow}\hl{changes} in the weather? \\
\midrule
\multicolumn{1}{c}{\textbf{Vanilla LLM}}  \\
\textbf{nDCG@10:} \textcolor{darkgray}{76.63\%} \\
\textbf{Expanded Query}: The \sethlcolor{lightgreen}\hl{coronavirus}, like many other \sethlcolor{lightgreen}\hl{viruses}, has been observed to respond to \sethlcolor{lightgreen}\hl{changes} in the weather by experiencing fluctuations in \sethlcolor{lightgreen}\hl{transmission} and spread. ... the \sethlcolor{lightgreen}\hl{virus} tends to thrive in environments with high \sethlcolor{lightgreen}\hl{humidity}, typically above 40\%, and a \sethlcolor{lightgreen}\hl{temperature} range of 37°C to 46°C. ... studies have found that the \sethlcolor{lightgreen}\hl{virus} can survive on surfaces for longer periods at lower \sethlcolor{lightgreen}\hl{temperatures} and \sethlcolor{lightgreen}\hl{humidity} levels, ... \\
\midrule

\multicolumn{1}{c}{\textbf{\method{}}} \\ 
\textbf{nDCG@10:} \textcolor{red}{100.00\%} \\
\textbf{Expanded Query}: The  \sethlcolor{lightblue}\hl{coronavirus} responds to \sethlcolor{lightblue}\hl{changes} in the weather by adapting its \sethlcolor{lightblue}\hl{transmission} and spread patterns. This is because \sethlcolor{lightblue}\hl{temperature}, \sethlcolor{lightblue}\hl{humidity}, and other \sethlcolor{lightblue}\hl{environmental factors} can affect the \sethlcolor{lightblue}\hl{stability} and survival of the \sethlcolor{lightblue}\hl{virus} on surfaces, ... research suggests that the \sethlcolor{lightblue}\hl{virus} may thrive in cooler and more humid environments, ... such as air circulation, ventilation, and \sethlcolor{lightblue}\hl{human behavior}. \\
\bottomrule
\end{tabular}
\caption{Case Study. All experiments are conducted based on the Contriever model under the zero-shot setting. To facilitate evaluation, we highlight the potential matching phrases between the golden passage and both the original and expanded queries. Different colors are used to annotate these matched phrases for each method: \sethlcolor{lightyellow}\hl{Green} for Direct Retrieval, \sethlcolor{lightgreen}\hl{Red} for Vanilla LLM, and \sethlcolor{lightblue}\hl{Blue} for \method{}.}
\label{tab:case_study}
\end{table*}

%% file: acl_latex.bbl
\begin{thebibliography}{58}
\providecommand{\natexlab}[1]{#1}

\bibitem[{Achiam et~al.(2023)Achiam, Adler, Agarwal, Ahmad, Akkaya, Aleman, Almeida, Altenschmidt, Altman, Anadkat et~al.}]{achiam2023gpt}
Josh Achiam, Steven Adler, Sandhini Agarwal, Lama Ahmad, Ilge Akkaya, Florencia~Leoni Aleman, Diogo Almeida, Janko Altenschmidt, Sam Altman, Shyamal Anadkat, et~al. 2023.
\newblock \href {https://arxiv.org/abs/2303.08774} {Gpt-4 technical report}.
\newblock \emph{ArXiv preprint}.

\bibitem[{AI@Meta(2024)}]{llama3modelcard}
AI@Meta. 2024.
\newblock \href {https://github.com/meta-llama/llama3/blob/main/MODEL_CARD.md} {Llama 3 model card}.

\bibitem[{Amini et~al.(2024)Amini, Vieira, and Cotterell}]{amini2024direct}
Afra Amini, Tim Vieira, and Ryan Cotterell. 2024.
\newblock \href {https://arxiv.org/abs/2402.10571} {Direct preference optimization with an offset}.
\newblock \emph{ArXiv preprint}.

\bibitem[{Bajaj et~al.(2016)Bajaj, Campos, Craswell, Deng, Gao, Liu, Majumder, McNamara, Mitra, Nguyen et~al.}]{bajaj2016ms}
Payal Bajaj, Daniel Campos, Nick Craswell, Li~Deng, Jianfeng Gao, Xiaodong Liu, Rangan Majumder, Andrew McNamara, Bhaskar Mitra, Tri Nguyen, et~al. 2016.
\newblock \href {https://ceur-ws.org/Vol-1773/CoCoNIPS\_2016\_paper9.pdf} {Ms marco: A human generated machine reading comprehension dataset}.
\newblock In \emph{Proceedings of {NeurIPS}}.

\bibitem[{Belkin et~al.(1982)Belkin, Oddy, and Brooks}]{belkin1982ask}
Nicholas~J Belkin, Robert~N Oddy, and Helen~M Brooks. 1982.
\newblock \href {https://www.emerald.com/insight/content/doi/10.1108/eb026722/full/html} {Ask for information retrieval: Part i. background and theory}.
\newblock \emph{Journal of documentation}, 38(2):61--71.

\bibitem[{Chen et~al.(2024)Chen, Xiao, Zhang, Luo, Lian, and Liu}]{chen2024bge}
Jianlv Chen, Shitao Xiao, Peitian Zhang, Kun Luo, Defu Lian, and Zheng Liu. 2024.
\newblock \href {https://doi.org/10.48550/arxiv.2402.03216} {Bge m3-embedding: Multi-lingual, multi-functionality, multi-granularity text embeddings through self-knowledge distillation}.
\newblock \emph{ArXiv preprint}.

\bibitem[{Devlin et~al.(2019)Devlin, Chang, Lee, and Toutanova}]{devlin2019bert}
Jacob Devlin, Ming-Wei Chang, Kenton Lee, and Kristina Toutanova. 2019.
\newblock \href {https://aclanthology.org/N19-1423} {{BERT}: Pre-training of deep bidirectional transformers for language understanding}.
\newblock In \emph{Proceedings of {NAACL-HLT}}, pages 4171--4186.

\bibitem[{Douze et~al.(2024)Douze, Guzhva, Deng, Johnson, Szilvasy, Mazaré, Lomeli, Hosseini, and Jégou}]{douze2024faiss}
Matthijs Douze, Alexandr Guzhva, Chengqi Deng, Jeff Johnson, Gergely Szilvasy, Pierre-Emmanuel Mazaré, Maria Lomeli, Lucas Hosseini, and Hervé Jégou. 2024.
\newblock \href {https://doi.org/10.48550/arxiv.2401.08281} {The faiss library}.
\newblock \emph{ArXiv preprint}.

\bibitem[{Fan et~al.(2019)Fan, Jernite, Perez, Grangier, Weston, and Auli}]{fan2019eli5}
Angela Fan, Yacine Jernite, Ethan Perez, David Grangier, Jason Weston, and Michael Auli. 2019.
\newblock \href {https://aclanthology.org/P19-1346} {{ELI}5: Long form question answering}.
\newblock In \emph{Proceedings of {ACL}}, pages 3558--3567.

\bibitem[{Fang et~al.(2024)Fang, Zhan, Ai, Mao, Su, Chen, and Liu}]{fang2024scaling}
Yan Fang, Jingtao Zhan, Qingyao Ai, Jiaxin Mao, Weihang Su, Jia Chen, and Yiqun Liu. 2024.
\newblock \href {https://dl.acm.org/doi/abs/10.1145/3626772.3657743} {Scaling laws for dense retrieval}.
\newblock In \emph{Proceedings of SIGIR}, pages 1339--1349.

\bibitem[{Gao and Callan(2021)}]{cocondenser}
Luyu Gao and Jamie Callan. 2021.
\newblock \href {https://aclanthology.org/2021.emnlp-main.75} {Condenser: a pre-training architecture for dense retrieval}.
\newblock In \emph{Proceedings of {EMNLP}}, pages 981--993.

\bibitem[{Gao and Callan(2022)}]{gao2022unsupervised}
Luyu Gao and Jamie Callan. 2022.
\newblock \href {https://aclanthology.org/2022.acl-long.203} {Unsupervised corpus aware language model pre-training for dense passage retrieval}.
\newblock In \emph{Proceedings of {ACL}}, pages 2843--2853.

\bibitem[{Gao et~al.(2023)Gao, Ma, Lin, and Callan}]{gao2023precise}
Luyu Gao, Xueguang Ma, Jimmy Lin, and Jamie Callan. 2023.
\newblock \href {https://aclanthology.org/2023.acl-long.99} {Precise zero-shot dense retrieval without relevance labels}.
\newblock In \emph{Proceedings of {ACL}}, pages 1762--1777.

\bibitem[{GLM et~al.(2024)GLM, Zeng, Xu, Wang, Zhang, Yin, Zhang, Rojas, Feng, Zhao et~al.}]{glm2024chatglm}
Team GLM, Aohan Zeng, Bin Xu, Bowen Wang, Chenhui Zhang, Da~Yin, Dan Zhang, Diego Rojas, Guanyu Feng, Hanlin Zhao, et~al. 2024.
\newblock \href {https://arxiv.org/abs/2406.12793} {Chatglm: A family of large language models from glm-130b to glm-4 all tools}.
\newblock \emph{ArXiv preprint}.

\bibitem[{Han et~al.(2023)Han, Liu, Liu, and Xiong}]{han2023enhancing}
Peixuan Han, Zhenghao Liu, Zhiyuan Liu, and Chenyan Xiong. 2023.
\newblock \href {https://arxiv.org/abs/2310.16605} {Enhancing dense retrievers' robustness with group-level reweighting}.
\newblock \emph{ArXiv preprint}.

\bibitem[{Hofst{\"a}tter et~al.(2021)Hofst{\"a}tter, Lin, Yang, Lin, and Hanbury}]{hofstatter2021efficiently}
Sebastian Hofst{\"a}tter, Sheng-Chieh Lin, Jheng-Hong Yang, Jimmy Lin, and Allan Hanbury. 2021.
\newblock \href {https://dl.acm.org/doi/abs/10.1145/3404835.3462891} {Efficiently teaching an effective dense retriever with balanced topic aware sampling}.
\newblock In \emph{Proceedings of SIGIR}, pages 113--122.

\bibitem[{Hu et~al.(2022)Hu, Shen, Wallis, Allen{-}Zhu, Li, Wang, Wang, and Chen}]{hu2022lora}
Edward~J. Hu, Yelong Shen, Phillip Wallis, Zeyuan Allen{-}Zhu, Yuanzhi Li, Shean Wang, Lu~Wang, and Weizhu Chen. 2022.
\newblock \href {https://openreview.net/forum?id=nZeVKeeFYf9} {Lora: Low-rank adaptation of large language models}.
\newblock In \emph{Proceedings of {ICLR}}.

\bibitem[{Huang et~al.(2024{\natexlab{a}})Huang, Mu, Wu, Li, Xiao, Xiao, and Zhu}]{huang2024translate}
Pengcheng Huang, Yongyu Mu, Yuzhang Wu, Bei Li, Chunyang Xiao, Tong Xiao, and Jingbo Zhu. 2024{\natexlab{a}}.
\newblock \href {https://link.springer.com/chapter/10.1007/978-981-97-8367-0_8} {Translate-and-revise: Boosting large language models for constrained translation}.
\newblock In \emph{China National Conference on Chinese Computational Linguistics}, pages 120--139.

\bibitem[{Huang et~al.(2024{\natexlab{b}})Huang, Zhang, Wu, Zhang, Wang, Wang, Cao, Xu, Zhao, Qin et~al.}]{huang2024scalingnote}
Suyuan Huang, Chao Zhang, Yuanyuan Wu, Haoxin Zhang, Yuan Wang, Maolin Wang, Shaosheng Cao, Tong Xu, Xiangyu Zhao, Zengchang Qin, et~al. 2024{\natexlab{b}}.
\newblock \href {https://doi.org/10.48550/arXiv.2411.15766} {Scalingnote: Scaling up retrievers with large language models for real-world dense retrieval}.
\newblock \emph{ArXiv preprint}.

\bibitem[{Ingwersen(1996)}]{ingwersen1996cognitive}
Peter Ingwersen. 1996.
\newblock \href {https://www.emerald.com/insight/content/doi/10.1108/eb026960/full/html} {Cognitive perspectives of information retrieval interaction: elements of a cognitive ir theory}.
\newblock \emph{Journal of documentation}, 52(1):3--50.

\bibitem[{Izacard et~al.(2021)Izacard, Caron, Hosseini, Riedel, Bojanowski, Joulin, and Grave}]{izacard2021unsupervised}
Gautier Izacard, Mathilde Caron, Lucas Hosseini, Sebastian Riedel, Piotr Bojanowski, Armand Joulin, and Edouard Grave. 2021.
\newblock \href {https://openreview.net/forum?id=jKN1pXi7b0} {Unsupervised dense information retrieval with contrastive learning}.
\newblock \emph{Transactions on Machine Learning Research (TMLR)}.

\bibitem[{Jagerman et~al.(2023)Jagerman, Zhuang, Qin, Wang, and Bendersky}]{jagerman2023query}
Rolf Jagerman, Honglei Zhuang, Zhen Qin, Xuanhui Wang, and Michael Bendersky. 2023.
\newblock \href {https://arxiv.org/abs/2305.03653} {Query expansion by prompting large language models}.
\newblock \emph{ArXiv preprint}.

\bibitem[{Jiang et~al.(2025)Jiang, Lin, Cao, Tian, Kang, Wang, Sun, and Han}]{jiang2025deepretrieval}
Pengcheng Jiang, Jiacheng Lin, Lang Cao, Runchu Tian, SeongKu Kang, Zifeng Wang, Jimeng Sun, and Jiawei Han. 2025.
\newblock \href {https://arxiv.org/abs/2503.00223} {Deepretrieval: Hacking real search engines and retrievers with large language models via reinforcement learning}.
\newblock \emph{ArXiv preprint}.

\bibitem[{Johnson et~al.(2019)Johnson, Douze, and J{\'e}gou}]{johnson2019billion}
Jeff Johnson, Matthijs Douze, and Herv{\'e} J{\'e}gou. 2019.
\newblock \href {https://ieeexplore.ieee.org/abstract/document/8733051} {Billion-scale similarity search with gpus}.
\newblock \emph{IEEE Transactions on Big Data}, 7(3):535--547.

\bibitem[{Karpukhin et~al.(2020)Karpukhin, Oguz, Min, Lewis, Wu, Edunov, Chen, and Yih}]{karpukhin2020dense}
Vladimir Karpukhin, Barlas Oguz, Sewon Min, Patrick Lewis, Ledell Wu, Sergey Edunov, Danqi Chen, and Wen-tau Yih. 2020.
\newblock \href {https://aclanthology.org/2020.emnlp-main.550} {Dense passage retrieval for open-domain question answering}.
\newblock In \emph{Proceedings of {EMNLP}}, pages 6769--6781.

\bibitem[{Khattab et~al.(2021)Khattab, Potts, and Zaharia}]{khattab2021relevance}
Omar Khattab, Christopher Potts, and Matei Zaharia. 2021.
\newblock \href {https://direct.mit.edu/tacl/article/doi/10.1162/tacl_a_00405/107205/Relevance-guided-Supervision-for-OpenQA-with} {Relevance-guided supervision for openqa with colbert}.
\newblock \emph{Transactions of the association for computational linguistics}, 9:929--944.

\bibitem[{Kim and Baek(2025)}]{kim2025syntriever}
Minsang Kim and Seungjun Baek. 2025.
\newblock \href {https://arxiv.org/abs/2502.03824} {Syntriever: How to train your retriever with synthetic data from llms}.
\newblock \emph{ArXiv preprint}.

\bibitem[{Li et~al.(2024)Li, Zhuang, Hui, Qin, Lin, Jagerman, Wang, and Bendersky}]{li2024can}
Minghan Li, Honglei Zhuang, Kai Hui, Zhen Qin, Jimmy Lin, Rolf Jagerman, Xuanhui Wang, and Michael Bendersky. 2024.
\newblock \href {https://doi.org/10.1145/3626772.3657979} {Can query expansion improve generalization of strong cross-encoder rankers?}
\newblock In \emph{Proceedings of {SIGIR}}, pages 2321--2326.

\bibitem[{Li et~al.(2021)Li, Liu, Xiong, and Liu}]{li2021more}
Yizhi Li, Zhenghao Liu, Chenyan Xiong, and Zhiyuan Liu. 2021.
\newblock \href {https://dl.acm.org/doi/10.1145/3471158.3472245} {More robust dense retrieval with contrastive dual learning}.
\newblock In \emph{Proceedings of SIGIR}, pages 287--296.

\bibitem[{Lin et~al.(2020)Lin, Yang, Nogueira, Tsai, Wang, and Lin}]{lin2020conversational}
Sheng-Chieh Lin, Jheng-Hong Yang, Rodrigo Nogueira, Ming-Feng Tsai, Chuan-Ju Wang, and Jimmy Lin. 2020.
\newblock \href {https://arxiv.org/abs/2004.01909} {Conversational question reformulation via sequence-to-sequence architectures and pretrained language models}.
\newblock \emph{ArXiv preprint}.

\bibitem[{Ma et~al.(2025)Ma, Lin, Oguz, Lin, Yih, and Chen}]{ma2025drama}
Xueguang Ma, Xi~Victoria Lin, Barlas Oguz, Jimmy Lin, Wen-tau Yih, and Xilun Chen. 2025.
\newblock \href {https://arxiv.org/abs/2502.18460} {Drama: Diverse augmentation from large language models to smaller dense retrievers}.
\newblock \emph{ArXiv preprint}.

\bibitem[{Mackie et~al.(2023)Mackie, Chatterjee, and Dalton}]{mackie2023generative}
Iain Mackie, Shubham Chatterjee, and Jeffrey Dalton. 2023.
\newblock \href {https://doi.org/10.1145/3539618.3591992} {Generative relevance feedback with large language models}.
\newblock In \emph{Proceedings of {SIGIR}}, pages 2026--2031.

\bibitem[{Mao et~al.(2024)Mao, Jiang, Chen, Li, Wang, Wang, Xie, Huang, Chen, and Zhang}]{mao2024rafe}
Shengyu Mao, Yong Jiang, Boli Chen, Xiao Li, Peng Wang, Xinyu Wang, Pengjun Xie, Fei Huang, Huajun Chen, and Ningyu Zhang. 2024.
\newblock \href {https://arxiv.org/abs/2405.14431} {Rafe: Ranking feedback improves query rewriting for rag}.
\newblock In \emph{Proceedings of {EMNLP} Findings}.

\bibitem[{Nussbaum et~al.(2024)Nussbaum, Morris, Duderstadt, and Mulyar}]{nussbaum2024nomic}
Zach Nussbaum, John~X Morris, Brandon Duderstadt, and Andriy Mulyar. 2024.
\newblock \href {https://arxiv.org/abs/2402.01613} {Nomic embed: Training a reproducible long context text embedder}.
\newblock \emph{ArXiv preprint}.

\bibitem[{Ren et~al.(2021)Ren, Qu, Liu, Zhao, She, Wu, Wang, and Wen}]{ren2021rocketqav2}
Ruiyang Ren, Yingqi Qu, Jing Liu, Wayne~Xin Zhao, Qiaoqiao She, Hua Wu, Haifeng Wang, and Ji{-}Rong Wen. 2021.
\newblock \href {https://doi.org/10.18653/v1/2021.emnlp-main.224} {Rocketqav2: {A} joint training method for dense passage retrieval and passage re-ranking}.
\newblock In \emph{Proceedings of EMNLP}, pages 2825--2835.

\bibitem[{Robertson et~al.(2009)Robertson, Zaragoza et~al.}]{robertson2009probabilistic}
Stephen Robertson, Hugo Zaragoza, et~al. 2009.
\newblock \href {https://doi.org/10.1561/1500000019} {The probabilistic relevance framework: Bm25 and beyond}.
\newblock \emph{Foundations and Trends{\textregistered} in Information Retrieval}, pages 333--389.

\bibitem[{Shi et~al.(2024)Shi, Min, Yasunaga, Seo, James, Lewis, Zettlemoyer, and Yih}]{shi2024replug}
Weijia Shi, Sewon Min, Michihiro Yasunaga, Minjoon Seo, Rich James, Mike Lewis, Luke Zettlemoyer, and Wen-tau Yih. 2024.
\newblock \href {https://doi.org/10.18653/v1/2024.naacl-long.463} {Replug: Retrieval-augmented black-box language models}.
\newblock In \emph{Proceedings of NAACL}, pages 8371--8384.

\bibitem[{Springer et~al.(2024)Springer, Kotha, Fried, Neubig, and Raghunathan}]{springer2024repetition}
Jacob~Mitchell Springer, Suhas Kotha, Daniel Fried, Graham Neubig, and Aditi Raghunathan. 2024.
\newblock \href {https://arxiv.org/abs/2402.15449} {Repetition improves language model embeddings}.
\newblock \emph{ArXiv preprint}.

\bibitem[{Thakur et~al.(2021)Thakur, Reimers, R{\"u}ckl{\'e}, Srivastava, and Gurevych}]{thakur2021beir}
Nandan Thakur, Nils Reimers, Andreas R{\"u}ckl{\'e}, Abhishek Srivastava, and Iryna Gurevych. 2021.
\newblock \href {https://arxiv.org/abs/2104.08663} {Beir: A heterogenous benchmark for zero-shot evaluation of information retrieval models}.
\newblock \emph{ArXiv preprint}.

\bibitem[{Thorne et~al.(2018)Thorne, Vlachos, Christodoulopoulos, and Mittal}]{thorne2018fever}
James Thorne, Andreas Vlachos, Christos Christodoulopoulos, and Arpit Mittal. 2018.
\newblock \href {https://aclanthology.org/N18-1074} {{FEVER}: a large-scale dataset for fact extraction and {VER}ification}.
\newblock In \emph{Proceedings of ACL}, pages 809--819.

\bibitem[{Trivedi et~al.(2023)Trivedi, Balasubramanian, Khot, and Sabharwal}]{trivedi2023interleaving}
Harsh Trivedi, Niranjan Balasubramanian, Tushar Khot, and Ashish Sabharwal. 2023.
\newblock \href {https://aclanthology.org/2023.acl-long.557} {Interleaving retrieval with chain-of-thought reasoning for knowledge-intensive multi-step questions}.
\newblock In \emph{Proceedings of {ACL}}, pages 10014--10037.

\bibitem[{Wang et~al.(2022)Wang, Yang, Huang, Jiao, Yang, Jiang, Majumder, and Wei}]{wang2022text}
Liang Wang, Nan Yang, Xiaolong Huang, Binxing Jiao, Linjun Yang, Daxin Jiang, Rangan Majumder, and Furu Wei. 2022.
\newblock \href {https://arxiv.org/abs/2212.03533} {Text embeddings by weakly-supervised contrastive pre-training}.
\newblock \emph{ArXiv preprint}.

\bibitem[{Wang et~al.(2024)Wang, Yang, Huang, Yang, Majumder, and Wei}]{wang2024improving}
Liang Wang, Nan Yang, Xiaolong Huang, Linjun Yang, Rangan Majumder, and Furu Wei. 2024.
\newblock \href {https://arxiv.org/abs/2401.00368} {Improving text embeddings with large language models}.
\newblock \emph{ArXiv preprint}.

\bibitem[{Wang et~al.(2023{\natexlab{a}})Wang, Yang, and Wei}]{wang2023query2doc}
Liang Wang, Nan Yang, and Furu Wei. 2023{\natexlab{a}}.
\newblock \href {https://aclanthology.org/2023.emnlp-main.585} {Query2doc: Query expansion with large language models}.
\newblock In \emph{Proceedings of {EMNLP}}, pages 9414--9423.

\bibitem[{Wang et~al.(2023{\natexlab{b}})Wang, MacAvaney, Macdonald, and Ounis}]{wang2023generative}
Xiao Wang, Sean MacAvaney, Craig Macdonald, and Iadh Ounis. 2023{\natexlab{b}}.
\newblock \href {https://arxiv.org/abs/2308.00415} {Generative query reformulation for effective adhoc search}.
\newblock \emph{ArXiv preprint}.

\bibitem[{Wei et~al.(2022{\natexlab{a}})Wei, Tay, Bommasani, Raffel, Zoph, Borgeaud, Yogatama, Bosma, Zhou, Metzler et~al.}]{wei2022emergent}
Jason Wei, Yi~Tay, Rishi Bommasani, Colin Raffel, Barret Zoph, Sebastian Borgeaud, Dani Yogatama, Maarten Bosma, Denny Zhou, Donald Metzler, et~al. 2022{\natexlab{a}}.
\newblock \href {https://openreview.net/forum?id=yzkSU5zdwD} {Emergent abilities of large language models}.
\newblock \emph{Trans. Mach. Learn. Res.}, 2022.

\bibitem[{Wei et~al.(2022{\natexlab{b}})Wei, Wang, Schuurmans, Bosma, Ichter, Xia, Chi, Le, and Zhou}]{wei2022chain}
Jason Wei, Xuezhi Wang, Dale Schuurmans, Maarten Bosma, Brian Ichter, Fei Xia, Ed~H. Chi, Quoc~V. Le, and Denny Zhou. 2022{\natexlab{b}}.
\newblock \href {http://papers.nips.cc/paper\_files/paper/2022/hash/9d5609613524ecf4f15af0f7b31abca4-Abstract-Conference.html} {Chain-of-thought prompting elicits reasoning in large language models}.
\newblock In \emph{Proceedings of {NeurIPS}}.

\bibitem[{Weller et~al.(2024)Weller, Lo, Wadden, Lawrie, Van~Durme, Cohan, and Soldaini}]{weller2024generative}
Orion Weller, Kyle Lo, David Wadden, Dawn Lawrie, Benjamin Van~Durme, Arman Cohan, and Luca Soldaini. 2024.
\newblock \href {https://aclanthology.org/2024.findings-eacl.134} {When do generative query and document expansions fail? a comprehensive study across methods, retrievers, and datasets}.
\newblock In \emph{Proceedings of EACL Findings}, pages 1987--2003.

\bibitem[{Xie et~al.(2023)Xie, Liu, and Xiong}]{xie2023unsupervised}
Yiqing Xie, Xiao Liu, and Chenyan Xiong. 2023.
\newblock \href {https://doi.org/10.1145/3539618.3592080} {Unsupervised dense retrieval training with web anchors}.
\newblock In \emph{Proceedings of {SIGIR}}, pages 2476--2480.

\bibitem[{Xiong et~al.(2021{\natexlab{a}})Xiong, Xiong, Li, Tang, Liu, Bennett, Ahmed, and Overwijk}]{xiong2021approximate}
Lee Xiong, Chenyan Xiong, Ye~Li, Kwok{-}Fung Tang, Jialin Liu, Paul~N. Bennett, Junaid Ahmed, and Arnold Overwijk. 2021{\natexlab{a}}.
\newblock \href {https://openreview.net/forum?id=zeFrfgyZln} {Approximate nearest neighbor negative contrastive learning for dense text retrieval}.
\newblock In \emph{Proceedings of {ICLR}}.

\bibitem[{Xiong et~al.(2021{\natexlab{b}})Xiong, Li, Iyer, Du, Lewis, Wang, Mehdad, Yih, Riedel, Kiela, and Oguz}]{xiong2021dense}
Wenhan Xiong, Xiang~Lorraine Li, Srini Iyer, Jingfei Du, Patrick S.~H. Lewis, William~Yang Wang, Yashar Mehdad, Scott Yih, Sebastian Riedel, Douwe Kiela, and Barlas Oguz. 2021{\natexlab{b}}.
\newblock \href {https://openreview.net/forum?id=EMHoBG0avc1} {Answering complex open-domain questions with multi-hop dense retrieval}.
\newblock In \emph{Proceedings of {ICLR}}.

\bibitem[{Yang et~al.(2024)Yang, Yang, Zhang, Hui, Zheng, Yu, Li, Liu, Huang, Wei et~al.}]{yang2024qwen2}
An~Yang, Baosong Yang, Beichen Zhang, Binyuan Hui, Bo~Zheng, Bowen Yu, Chengyuan Li, Dayiheng Liu, Fei Huang, Haoran Wei, et~al. 2024.
\newblock \href {https://arxiv.org/abs/2412.15115} {Qwen2. 5 technical report}.
\newblock \emph{ArXiv preprint}.

\bibitem[{Yang et~al.(2018)Yang, Qi, Zhang, Bengio, Cohen, Salakhutdinov, and Manning}]{yang2018hotpotqa}
Zhilin Yang, Peng Qi, Saizheng Zhang, Yoshua Bengio, William Cohen, Ruslan Salakhutdinov, and Christopher~D. Manning. 2018.
\newblock \href {https://aclanthology.org/D18-1259} {{H}otpot{QA}: A dataset for diverse, explainable multi-hop question answering}.
\newblock In \emph{Proceedings of {EMNLP}}, pages 2369--2380.

\bibitem[{Ye et~al.(2023)Ye, Fang, Li, and Yilmaz}]{ye2023enhancing}
Fanghua Ye, Meng Fang, Shenghui Li, and Emine Yilmaz. 2023.
\newblock \href {https://doi.org/10.18653/v1/2023.findings-emnlp.398} {Enhancing conversational search: Large language model-aided informative query rewriting}.
\newblock In \emph{Proceedings of EMNLP Findings}, pages 5985--6006.

\bibitem[{Yu et~al.(2020)Yu, Liu, Yang, Xiong, Bennett, Gao, and Liu}]{yu2020few}
Shi Yu, Jiahua Liu, Jingqin Yang, Chenyan Xiong, Paul Bennett, Jianfeng Gao, and Zhiyuan Liu. 2020.
\newblock \href {https://doi.org/10.1145/3397271.3401323} {Few-shot generative conversational query rewriting}.
\newblock In \emph{Proceedings of SIGIR}, pages 1933--1936.

\bibitem[{Yu et~al.(2021)Yu, Liu, Xiong, Feng, and Liu}]{Yu2021FewShotCD}
Shi Yu, Zhenghao Liu, Chenyan Xiong, Tao Feng, and Zhiyuan Liu. 2021.
\newblock \href {https://doi.org/10.1145/3404835.3462856} {Few-shot conversational dense retrieval}.
\newblock In \emph{Proceedings of {SIGIR}}, pages 829--838.

\bibitem[{Zhan et~al.(2021)Zhan, Mao, Liu, Guo, Zhang, and Ma}]{zhan2021optimizing}
Jingtao Zhan, Jiaxin Mao, Yiqun Liu, Jiafeng Guo, Min Zhang, and Shaoping Ma. 2021.
\newblock \href {https://doi.org/10.1145/3404835.3462880} {Optimizing dense retrieval model training with hard negatives}.
\newblock In \emph{Proceedings of {SIGIR}}, pages 1503--1512.

\bibitem[{Zhang et~al.(2025)Zhang, Bi, Guo, Sun, Liu, Shi, Yin, and Cheng}]{zhang2025unleashing}
Hengran Zhang, Keping Bi, Jiafeng Guo, Xiaojie Sun, Shihao Liu, Daiting Shi, Dawei Yin, and Xueqi Cheng. 2025.
\newblock \href {https://arxiv.org/abs/2504.05216} {Unleashing the power of llms in dense retrieval with query likelihood modeling}.
\newblock \emph{ArXiv preprint}.

\end{thebibliography}
